\documentclass{amsart}
\usepackage{amsfonts}

\usepackage{graphicx}

\input{tcilatex}
\input tcilatex

\begin{document}
\title[Gibbs-Poisson sampling and occupancies]{Occupancy distributions arising in sampling from Gibbs-Poisson abundance
models}
\author{Thierry Huillet$^{1}$, Servet Mart\'{i}nez$^{2}$}
\address{$^{1}$Laboratoire de Physique Th\'{e}orique et Mod\'{e}lisation\\
Universit\'{e} de Cergy-Pontoise\\
CNRS UMR-8089\\
Site de Saint Martin\\
2 avenue Adolphe-Chauvin\\
95302 Cergy-Pontoise, France\\
$^{2}$Depto. Ingenieria Matematica and Centro Modelamiento Matematico\\
Universidad de Chile\\
UMI 2071, Uchile-Cnrs\\
Casilla 170-3 Correo 3\\
Santiago, Chile\\
\\
E-mail: huillet@u-cergy.fr, smartine@dim.uchile.cl}
\maketitle

\begin{abstract}
Estimating the number $n$ of unseen species from a $k-$sample displaying
only $p\leq k$ distinct sampled species has received attention for long. It
requires a model of species abundance together with a sampling model. We
start with a discrete model of iid stochastic species abundances, each with
Gibbs-Poisson distribution. A $k-$sample drawn from the $n-$species
abundances vector is the one obtained while conditioning it on summing to $k$%
. We discuss the sampling formulae (species occupancy distributions,
frequency of frequencies) in this context. We then develop some aspects of
the estimation of $n$ problem from the size $k$ of the sample and the
observed value of $P_{n,k}$, the number of distinct sampled species.

It is shown that it always makes sense to study these occupancy problems
from a Gibbs-Poisson abundance model in the context of a population with
infinitely many species. From this extension, a parameter $\gamma $
naturally appears, which is a measure of richness or diversity of species.
We rederive the sampling formulae for a population with infinitely many
species, together with the distribution of the number $P_{k}$ of distinct
sampled species. We investigate the estimation of $\gamma $ problem from the
sample size $k$ and the observed value of $P_{k}$.

We then exhibit a large special class of Gibbs-Poisson distributions having
the property that sampling from a discrete abundance model may equivalently
be viewed as a sampling problem from a random partition of unity, now in the
continuum. When $n$ is finite, this partition may be built upon normalizing $%
n$ infinitely divisible iid positive random variables by its partial sum. It
is shown that the sampling process in the continuum should generically be
biased on the total length appearing in the latter normalization. A
construction with size-biased sampling from the ranked normalized jumps of a
subordinator is also supplied, would the problem under study present
infinitely many species. We illustrate our point of view with many examples,
some of which being new ones.\newline


\textbf{Running title}: Gibbs-Poisson sampling and occupancies.
\end{abstract}

\section{Introduction and outline of main results}

Estimating the number $n$ of unseen species from a $k-$sample displaying
only $p\leq k$ distinct sampled species has been a challenging problem since
the mid-twentieth century, \cite{FCW}. It requires a model of species
abundance together with a sampling model \cite{Engen}, and the answer to the
latter question is of course model-dependent. In this work, we start with a
discrete model of independent and identically distributed (iid) stochastic
species abundances $\mathbf{\xi }_{n}:=\left( \xi _{1},...,\xi _{n}\right) ,$
based on compound Poisson distributions for $\xi \overset{d}{=}\xi _{1}$. We
discuss the sampling formulae (species occupancy distributions, frequency of
frequencies) in this discrete context. Typically, a $k-$sample drawn from
the $n-$species abundances vector is the one obtained while conditioning
this vector on summing to $k$ (the sample size). Sampling from iid compound
Poisson abundance random variables (rvs) in this sense results in a
Gibbs-Poisson sampling model from $\mathbf{\xi }_{n}$. It has to do with
random allocation of balls into boxes, \cite{Kol}, \cite{Kol2}. Various
combinatorial identities arising in this setup are discussed. A distribution
for the number of distinct visited species $P_{n,k}$ in a $k-$sample from a
population of size $n$ with compound Poisson abundance is derived. For this
class of sampling problems, a `temperature' type parameter $\theta >0$ pops
in naturally. It is a measure of how similar the box occupancy numbers look
like statistically, after the sampling process: the smaller the values of $%
\theta $, the more likely it is that these occupancy numbers are disparate.
When sampling from $\mathbf{\xi }_{n}$, we then discuss some aspects of the
problem of the estimation of the number of species $n$ from the size $k$ of
the sample and the number $P_{n,k}$ of distinct sampled species, assuming $%
\theta $ to be known. These results are supplied in Propositions $1$ and $3$.

It turns out that it always makes sense to study these occupancy problems
from a Gibbs-Poisson abundance model in the context of a population with
infinitely many species, provided $n$ goes to $\infty $ together with $%
\theta $ going to $0$ while $n\theta \rightarrow \gamma >0$. From this
construction, $\gamma $ then appears as a measure of species richness or
diversity. We rederive the sampling formulae (species occupancy
distributions, frequency of frequencies) for a population with infinitely
many species, together with the distribution of the number $P_{k}$ of
distinct sampled species. We discuss the problem of the estimation of the
diversity parameter $\gamma $ from the size $k$ of the sample and the number 
$P_{k}$.

One particular model in the compound Poisson class has been discussed at
length in the literature: the sampling problem from a population with
discrete negative binomial distribution abundance $\xi $, both when the
population is made of a finite number of species and when there are
infinitely many of them. For this particular model, when there are
infinitely many species, the obtained sampling formulae are the ones of
Ewens, \cite{Ewen}. It is also well-known that the Ewens sampling formulae
may also be viewed as sampling from a random Dirichlet partition of the
unity when the number of species is finite or as sampling from a random
Poisson-Dirichlet partition of unity when there are infinitely many classes, 
\cite{Holst}. This property is remarkable. By sampling from a partition of
the continuum $\left[ 0,1\right] $, we mean that we draw independently $k$
uniform random variables on the unit interval, looking at the subintervals
of the partition which are being hit in the process to form the occupancy
distributions of classes.

In this work, we exhibit a large class of compound Poisson distributions
sharing with the negative binomial distribution this property that sampling
from a discrete abundance model may equivalently be viewed as a sampling
problem from a random partition of unity in the continuum. When $n$ is
finite, this partition may be built upon normalizing $n$ infinitely
divisible iid non-negative random variables $\mathbf{Y}_{n}:=\left(
Y_{1},Y_{2},...,Y_{n}\right) $ by its partial sum. We exhibit the one-to-one
correspondence between the laws of $\xi $ and $Y\overset{d}{=}Y_{1}$,
assuming $\xi $ to be in the special class. It is however shown that the
sampling process in the continuum should generically be biased on the total
length appearing in the latter normalization. A construction with such
biased sampling from the ranked normalized jumps of a subordinator is also
supplied, would the problem under study present infinitely many species. The
biasing factors account for the fact that the Gibbs-Poisson occupancy models
are not in general sampling consistent as $k$ varies (are not EPPFs). A
complete classification of EPPFs induced by the unbiased multinomial
sampling from partition of unity can be found in \cite{GP}, \cite{Ho}.

With this correspondence in mind, we discuss several examples, among which
the Engen extended negative binomial model \cite{Eng}, the Berestycki-Pitman
model \cite{Beres} for the enumeration of forests of trees with generalized
binomial generator, the polylog and the Mittag-Leffler models. When there
are some reasons to suspect that the ranked species frequencies decay
algebraically with the rank number, then the Engen model is well suited.
Would one think of the ranked species frequencies as decaying exponentially
with the rank number, then the Ewens model seems relevant. If the ranked
species frequencies are believed to decay exponentially as some power of the
rank number, then one should opt for the polylog model.

We end up giving a new example of $\xi $ sharing some common issues with the
Engen's model (in particular the algebraic decay property of the ranked
frequencies). For this precise model, we are able to give an exact estimator
of the biodiversity parameter.

\section{Sampling from discrete Gibbs-Poisson distributions}

The sampling problem from a negative binomial abundance model and its
Dirichlet counterpart in the continuum suggest to study the following
general construction (see \cite{Hosh1}, \cite{Hosh2}, \cite{Beres}, \cite
{Kol} and \cite{PitCSP} for similar recent interest).

\subsection{Generating and partition function (see \protect\cite{Comtet} and 
\protect\cite{PitCSP}, Section $1$)}

With $\phi _{\bullet }:=\left( \phi _{m};m\geq 1\right) $ a sequence of
non-negative real numbers with $\phi _{1}>0$, let 
\begin{equation}
\phi \left( x\right) :=\sum_{m\geq 1}\frac{\phi _{m}}{m!}x^{m}  \label{f0}
\end{equation}
be a formal power series in $x$. Assume that $x_{0}:=\sup \left( x>0:\phi
\left( x\right) <\infty \right) \in \left( 0,+\infty \right] $ is its
convergence radius. Then $\phi \left( x\right) $ defines a convergent series
on $\left| x\right| <x_{0}$ and it is absolutely monotone on $\left(
0,x_{0}\right) $ in the sense that $\phi ^{\left( n\right) }\left( x\right)
\geq 0$ for all $n\geq 0$ and $x\in \left( 0,x_{0}\right) .$ We call it the
local exponential generating function.\newline

Let $\theta >0$ and consider the exponential `partition' generating function 
\begin{equation}
Z_{\theta }\left( x\right) =e^{\theta \phi \left( x\right) }.  \label{f1}
\end{equation}
This function also defines a convergent series on $\left| x\right| <x_{0}$
with $Z_{\theta }\left( 0\right) =1.$ Further, with $\sigma _{k}\left(
\theta \right) =k!\left[ x^{k}\right] Z_{\theta }\left( x\right) $ (where $%
\left[ x^{k}\right] f\left( x\right) $ is the $x^{k}-$coefficient in the
series expansion of the function $f\left( x\right) $): 
\begin{equation*}
Z_{\theta }\left( x\right) =1+\sum_{k\geq 1}\frac{x^{k}}{k!}\sigma
_{k}\left( \theta \right) .
\end{equation*}
Since $\partial _{x}Z_{\theta }\left( x\right) =\theta \phi ^{\prime }\left(
x\right) Z_{\theta }\left( x\right) $, we get the recurrence: 
\begin{equation}
\sigma _{k+1}\left( \theta \right) =\theta \sum_{l=0}^{k}\binom{k}{l}\phi
_{k-l+1}\sigma _{l}\left( \theta \right) \text{, }k\geq 0\text{, }\sigma
_{0}\left( \theta \right) \equiv 1.  \label{f2}
\end{equation}
Similarly, since $\partial _{\theta }Z_{\theta }\left( x\right) =:Z_{\theta
}^{\prime }\left( x\right) =\phi \left( x\right) Z_{\theta }\left( x\right) $%
, we find: 
\begin{equation}
\sigma _{k}^{\prime }\left( \theta \right) =\sum_{l=0}^{k-1}\binom{k}{l}\phi
_{k-l}\sigma _{l}\left( \theta \right) \text{, }k\geq 1\text{, }\sigma
_{0}\left( \theta \right) =1.  \label{f2b}
\end{equation}
Then, clearly, 
\begin{equation}
\sigma _{k}\left( \theta \right) =\sum_{l=1}^{k}B_{k,l}\left( \phi _{\bullet
}\right) \theta ^{l},  \label{f2a}
\end{equation}
with: 
\begin{equation*}
B_{k,l}\left( \phi _{\bullet }\right) =\frac{k!}{l!}\left[ x^{k}\right] \phi
\left( x\right) ^{l}=\frac{k!}{l!}\sum_{\mathbf{m}_{l}:\text{ }\left| 
\mathbf{m}_{l}\right| =k}\prod_{j=1}^{l}\frac{\phi _{m_{j}}}{m_{j}!}\geq 0.
\end{equation*}
In the latter sum, summation runs over $\mathbf{m}_{l}:=\left(
m_{1},...,m_{l}\right) \in \Bbb{N}^{l},$ with $\left| \mathbf{m}_{l}\right|
:=\sum_{j=1}^{l}m_{j}=$ $k$ and $\Bbb{N}:=\left\{ 1,2,...\right\} $; there
are $\binom{k-1}{l-1}$ terms in such sums. So $\sigma _{k}\left( \theta
\right) $ is a degree-$k$ Bell polynomial in $\theta $ whose $\theta ^{l}$
coefficient is $B_{k,l}\left( \phi _{\bullet }\right) $ which is known as
the Bell exponential polynomial in the variables $\phi _{\bullet }$ (see 
\cite{Comtet}). On $\theta >0$, the function $\sigma _{k}\left( \theta
\right) $ is convex and log-concave, for all $k$. As a polynomial with
non-negative coefficients of degree $k$, $\sigma _{k}\left( \theta \right) $
has no strictly positive real root and at most $k$ real non-positive roots
(including $0$), counting roots with their multiplicity.\newline

\textbf{Remarks }(Bell polynomials and convolutions).

$\left( i\right) $ Define $\left( \phi *\phi \right) _{m}:=\sum_{l=1}^{m-1}%
\binom{m}{l}\phi _{l}\phi _{m-l}$, $m\geq 2,$ as the binomial
self-convolution sequence of $\phi _{m}$. Define $\phi _{m}^{*p}$ as the $m^{%
\text{th}}$ term, $m\geq p,$ of the sequence $\phi ^{*p}:=\phi *...*\phi $, $%
p$ times; then the following convolution identity is well-known to hold: 
\begin{equation*}
B_{k,p}\left( \phi _{\bullet }\right) =\phi _{k}^{*p}/p!.
\end{equation*}

$\left( ii\right) $ Because $Z_{\theta +\theta ^{\prime }}\left( x\right)
=Z_{\theta }\left( x\right) Z_{\theta ^{\prime }}\left( x\right) $, the
polynomials $\sigma _{k}\left( \theta \right) $ satisfy 
\begin{equation}
\sigma _{k}\left( \theta +\theta ^{\prime }\right) =\sum_{l=0}^{k}\binom{k}{l%
}\sigma _{l}\left( \theta \right) \sigma _{k-l}\left( \theta ^{\prime
}\right) \text{ for all }\theta ,\theta ^{\prime }>0,  \label{f2c}
\end{equation}
and so they form a so-called binomial convolution sequence of polynomials.

If $p\geq 1$ is an integer, with $\sigma \left( 1\right) _{k}^{*p}:=\left(
\sigma \left( 1\right) ^{*p}\right) _{k},$ $\mathbf{k}_{p}:=\left(
k_{1},...,k_{p}\right) $ in $\Bbb{N}_{0}^{p},$ $\left| \mathbf{k}_{p}\right|
:=k_{1}+...+k_{p}$ and $\Bbb{N}_{0}:=\left\{ 0,1,2,...\right\} $%
\begin{equation*}
\sigma _{k}\left( p\right) =\sigma \left( 1\right) _{k}^{*p}=\sum_{\mathbf{k}%
_{p}\in \Bbb{N}_{0}^{p}:\text{ }\left| \mathbf{k}_{p}\right| =k}\binom{k}{%
k_{1}...k_{p}}\prod_{q=1}^{p}\sigma _{k_{q}}\left( 1\right) .
\end{equation*}
We clearly have 
\begin{equation*}
\sigma _{k}\left( p\right) =\sum_{q=1}^{p}\binom{p}{q}\sum_{\mathbf{k}%
_{q}\in \Bbb{N}^{q}:\text{ }\left| \mathbf{k}_{p}\right| =k}\binom{k}{%
k_{1}...k_{q}}\prod_{r=1}^{q}\sigma _{k_{r}}\left( 1\right) .
\end{equation*}
In other words, 
\begin{equation}
\sigma _{k}\left( p\right) =\sum_{q=1}^{k}\binom{p}{q}\sum_{\mathbf{k}%
_{q}\in \Bbb{N}^{q}:\text{ }\left| \mathbf{k}_{p}\right| =k}\binom{k}{%
k_{1}...k_{q}}\prod_{r=1}^{q}\sigma _{k_{r}}\left( 1\right) ,  \label{f1aa}
\end{equation}
where $\binom{p}{q}=0$ if $q>p.$ This expression extends to non-integral
arguments $\theta >0$ of $\sigma _{k}\left( \cdot \right) $ as 
\begin{equation}
\sigma _{k}\left( \theta \right) =:\sigma \left( 1\right) _{k}^{*\theta
}=\sum_{q=1}^{k}\binom{\theta }{q}\sum_{\mathbf{k}_{q}\in \Bbb{N}^{q}:\text{ 
}\left| \mathbf{k}_{p}\right| =k}\binom{k}{k_{1}...k_{q}}\prod_{r=1}^{q}%
\sigma _{k_{r}}\left( 1\right) ,  \label{f1ab}
\end{equation}
where $\binom{\theta }{q}=:\left\{ \theta \right\} _{q}/q!$ with $\left\{
\theta \right\} _{q}:=\Gamma \left( \theta +1\right) /\Gamma \left( \theta
-q+1\right) =\theta \left( \theta -1\right) ..\left( \theta -q+1\right) ,$
the usual extension of $\binom{p}{q}$ for the expansion of $\left(
1+x\right) ^{\theta }.$ From (\ref{f1ab}), it is clear again that $\sigma
_{k}\left( \theta \right) $ is a degree$-k$ polynomial in $\theta $ with no
constant term. This expression should be used instead of (\ref{f2a})
whenever the values at $\theta =1$ of $\sigma _{k}\left( \cdot \right) $ are
available in the first place, instead of the $\phi _{\bullet }$.\newline

$\left( iii\right) $ Putting the expression of $\sigma _{k}\left( \theta
\right) $ in (\ref{f2a}) into the recurrence equation (\ref{f2}) which $%
\left( \sigma _{k}\left( \theta \right) ;k\geq 1\right) $ satisfies gives 
\begin{equation}
l\cdot B_{k,l}\left( \phi _{\bullet }\right) =\sum_{j=l-1}^{k-1}\binom{k}{j}%
\phi _{k-j}B_{j,l-1}\left( \phi _{\bullet }\right) \text{.}  \label{f3}
\end{equation}
Recalling the boundary conditions 
\begin{equation*}
B_{k,0}\left( \phi _{\bullet }\right) =B_{0,l}\left( \phi _{\bullet }\right)
=0,\text{ }k,l\geq 1\text{ and }B_{0,0}\left( \phi _{\bullet }\right) :=1,
\end{equation*}
we get 
\begin{equation}
B_{k,1}\left( \phi _{\bullet }\right) =\phi _{k}\text{ and }B_{k,k}\left(
\phi _{\bullet }\right) =\phi _{1}^{k}\text{.}  \label{f4}
\end{equation}

$\left( iv\right) $ While performing the substitution $\theta \rightarrow
1/\theta $, $\sigma _{k}\left( \theta \right) $ should be mapped into the
new polynomial with respect to $1/\theta $%
\begin{equation*}
\sigma _{k}\left( 1/\theta \right) =\theta ^{-\left( k+1\right)
}\sum_{l=1}^{k}B_{k,k-l+1}\left( \phi _{\bullet }\right) \theta ^{l},
\end{equation*}
involving the `reversed' Bell sequence $B_{k,k-l+1}\left( \phi _{\bullet
}\right) .$

\subsection{Discrete compound Poisson distributions arising from $Z_{\theta
}\left( x\right) $}

Let now $\xi \in \Bbb{N}_{0}$ be a discrete random variable whose
probability generating (pgf) is given by: 
\begin{equation*}
\Phi \left( u\right) :=\mathbf{E}\left[ u^{\xi }\right] =\frac{Z_{\theta
}\left( xu\right) }{Z_{\theta }\left( x\right) },\text{ }\left| u\right|
\leq 1.
\end{equation*}
Since 
\begin{equation}
\mathbf{E}\left[ u^{\xi }\right] =e^{-\theta \phi \left( x\right) \left( 1-%
\frac{\phi \left( xu\right) }{\phi \left( x\right) }\right) },  \label{f5}
\end{equation}
$\xi $ is in the compound Poisson (CP) class, as a Poisson sum of iid jumps,
hence infinitely divisible$.$ The jumps' height law is given by its pgf $%
\mathbf{E}\left[ u^{\delta }\right] =\frac{\phi \left( xu\right) }{\phi
\left( x\right) },$ where $\delta \in \Bbb{N}$ is one of these jumps$.$ Note
that both $\mathbf{E}\left[ \delta \right] =x\frac{\phi ^{\prime }\left(
x\right) }{\phi \left( x\right) }$ and $\mathbf{E}\left[ \xi \right] =\theta
\phi \left( x\right) \mathbf{E}\left[ \delta \right] =\theta x\phi ^{\prime
}\left( x\right) $ are finite when $\left| x\right| <x_{0}$. Clearly 
\begin{equation*}
\mathbf{P}\left( \delta =m\right) =\frac{\phi _{m}x^{m}}{\phi \left(
x\right) \cdot m!},\text{ }m\geq 1\text{ and}
\end{equation*}
\begin{equation*}
\mathbf{P}\left( \xi =k\right) =\frac{\sigma _{k}\left( \theta \right) x^{k}%
}{Z_{\theta }\left( x\right) \cdot k!}\text{, }k\geq 0.
\end{equation*}
With $y$ defined by $x=:e^{-y}$, $y$ is indeed the Legendre conjugate of $%
\mu :=\mathbf{E}\left( \xi \right) $. So the parameter $x$ in (\ref{f5}) can
serve to adjust the mean $\mu $ of $\xi $. The random variable $\xi $ will
be used in the sequel as the abundance of some species in a population with $%
n$ species. Due to its compound Poisson structure, it is tacitly assumed
that species abundance is modelled as a Poisson sum of iid `clusters' each
with random size distributed like $\delta \geq 1.$\newline

Consider now a sequence $\mathbf{\xi }:=\left( \xi _{1},...,\xi
_{n},...\right) $ of iid compound Poisson random variables, each on $\Bbb{N}%
_{0}$. Let $\zeta _{n}:=\sum_{m=1}^{n}\xi _{m}$ denote their partial sum.
Then, because $\xi $ is in the compound-Poisson class due to $Z_{\theta
}\left( x\right) ^{n}=Z_{n\theta }\left( x\right) $%
\begin{equation*}
\mathbf{P}\left( \zeta _{n}=k\right) =\frac{\sigma _{k}\left( n\theta
\right) x^{k}}{Z_{n\theta }\left( x\right) \cdot k!}\text{, }k\geq 0.
\end{equation*}
This is also a compound Poisson distribution with corresponding partition
function $Z_{n\theta }\left( x\right) .$\newline

\textbf{Remark}: One could think of starting with $\phi \left( x\right)
:=\phi _{0}+\sum_{m\geq 1}\frac{\phi _{m}}{m!}x^{m}$ with $\phi _{0}\geq 0$
but because we shall deal with CP distributions whose pgfs are given by (\ref
{f5}), $\phi _{0}$ plays no role in our problem$.$

\subsection{Sampling from infinitely divisible CP distributions}

Define a random allocation scheme of $k$ distinguishable particles or balls
into $n$ distinguishable boxes by 
\begin{equation*}
\mathbf{K}_{n,k}:=\left( K_{n,k}\left( 1\right) ,...,K_{n,k}\left( n\right)
\right) \overset{d}{=}\left( \xi _{1},...,\xi _{n}\mid \zeta _{n}=k\right) ,
\end{equation*}
so that $K_{n,k}\left( m\right) $ counts the number of particles in box $m$, 
$m=1,...,n$ in a $k-$sample$.$ Defining $\mathbf{K}_{n,k}$ from $n$ iid $\xi 
$'s conditioned on summing to $k$, we get the generalized allocation scheme
defined by Kolchin, (see \cite{Kol}). When the $\xi $'s are in addition CP
distributed, we call this model sampling from Gibbs-Poisson (GP)
distributions. \newline

\textbf{Remark:} Since $\mathbf{E}\left[ \xi \right] =\Phi ^{\prime }\left(
1\right) =\theta x\phi ^{\prime }\left( x\right) ,$ $\theta >0$ and $x\in
\left( 0,x_{0}\right) $, we could adjust the mean $\mu $ of $\xi $ so that $%
\mathbf{E}\left[ \xi \right] =\mu .$ Then we would have the relation $\mu
/\theta =x\phi ^{\prime }\left( x\right) $ (Legendre conjugation of $x$ and $%
\mu $) from which, by Lagrange inversion formula, an expression of $x$ as a
function of $\mu /\theta $ would follow. However, as we shall see, the
actual value of the mean $\mu $ does not really matter after the sampling
process.\newline

Taking now into account the conditioning on the sample size in the
definition of $\mathbf{K}_{n,k}$'s law, with $\mathbf{k}_{n}:=\left(
k_{1},...,k_{n}\right) \in \Bbb{N}_{0}^{n}$ a vector of non-negative
integers obeying $\left| \mathbf{k}_{n}\right| :=\sum_{m=1}^{n}k_{m}=$ $k$%
\begin{equation}
\mathbf{P}\left( \mathbf{K}_{n,k}=\mathbf{k}_{n}\right) =\frac{\mathbf{P}%
\left( \xi _{1}=k_{1},...,\xi _{n}=k_{n}\right) }{\mathbf{P}\left( \zeta
_{n}=k\right) }=\frac{1}{\sigma _{k}\left( n\theta \right) }\binom{k}{%
k_{1}...k_{n}}\prod_{m=1}^{n}\sigma _{k_{m}}\left( \theta \right) ,
\label{f6}
\end{equation}
this (Maxwell-Boltzmann) joint law being independent of $x$ and so of the
mean $\mu $ of the $\xi $'s$.$ In other words, the joint probability
generating function of $\mathbf{K}_{n,k}$ reads ($\left| u_{m}\right| \leq
1; $ $m=1,...,n$): 
\begin{equation}
\mathbf{E}\left[ \prod_{m=1}^{n}u_{m}^{K_{n,k}\left( m\right) }\right] =%
\frac{1}{\sigma _{k}\left( n\theta \right) }\sum_{\mathbf{k}_{n}\in \Bbb{N}%
_{0}^{n}:\text{ }\left| \mathbf{k}_{n}\right| =k}\binom{k}{k_{1}...k_{n}}%
\prod_{m=1}^{n}\sigma _{k_{m}}\left( \theta \right) u_{m}^{k_{m}}.
\label{f7}
\end{equation}
From (\ref{f6}), $w_{k_{m}}\left( \theta \right) :=\sigma _{k_{m}}\left(
\theta \right) /k_{m}!$ is seen to be the Boltzmann weight of box $m$ with $%
e_{k_{m}}\left( \theta \right) :=-\log \left( \sigma _{k_{m}}\left( \theta
\right) /k_{m}!\right) $ being the energy required to put $k_{m}$ balls into
box number $m.$ More precisely, for our random allocation GP model of
particles (\ref{f7}) and from (\ref{f2a}), the price to pay for having the $%
l^{\text{th}}$ particle, $l\in \left\{ 1,...,k_{m}\right\} ,$ in box $m$
simply is $l$ and this event is assigned the weight $B_{k_{m},l}\left( \phi
_{\bullet }\right) /k_{m}!$. From this, one may view $\theta $ as a box
temperature parameter which, under our assumptions, is here common to all
boxes (or species). Due to $\sigma _{k_{m}}\left( \theta \right) $ being a
polynomial in $\theta $ with positive coefficients, the energy $%
e_{k_{m}}\left( \theta \right) $ is indeed a decreasing function of $\theta $
and one may therefore interpret $\theta $ as some temperature(\footnote{%
In statistical contexts, this temperature parameter is also called the
concentration parameter.}) of the boxes (maybe through the monotone
transformation $\theta \leftrightarrow e^{-1/T}$). Note that when $\theta $
approaches $0$, the energy $e_{k_{m}}\left( \theta \right) \sim -\log \theta 
$ tends to $+\infty :$ because the price to pay to put any number of
particles into a box is extremely high, the optimal strategy is to put them
all into a single box. One therefore expects that, as $\theta $ gets very
small, the vector $\mathbf{K}_{n,k}$ gets very skewed (most balls into a
single box), that is, completely opposite to the balanced multinomial$\left(
k;\frac{1}{n},...,\frac{1}{n}\right) $ situation 
\begin{equation*}
\mathbf{P}\left( \mathbf{K}_{n,k}=\mathbf{k}_{n}\right) =\frac{k!}{%
\prod_{m=1}^{n}k_{m}!}n^{-k},\text{ }\left| \mathbf{k}_{n}\right| =k,
\end{equation*}
which is obtained for $\theta \rightarrow \infty $, as a result of $\sigma
_{k_{m}}\left( \theta \right) \sim \left( \phi _{1}\theta \right) ^{k_{m}}$.
As a conclusion, smaller the values of $\theta $, the more likely it is that
the occupancy numbers $K_{n,k}\left( m\right) $ are disparate.\newline

From (\ref{f6}), the random vector-count $\mathbf{K}_{n,k}$ has exchangeable
distribution (invariance under any permutation of the boxes numbers). But
obviously, in the ordered version $\mathbf{K}_{\left( n\right) ,k}$ of the
box occupancies $\mathbf{K}_{n,k}$, say with $K_{\left( n\right) ,k}\left(
1\right) \geq ...\geq K_{\left( n\right) ,k}\left( n\right) ,$ the boxes are
not equally filled and so $\mathbf{K}_{\left( n\right) ,k}$ is not
exchangeable.\newline

- Let us now compute the distribution of one of its typical components, say $%
K_{n,k}\left( 1\right) $. With $l\in \left\{ 0,...,k\right\} $, we get 
\begin{equation*}
\mathbf{P}\left( K_{n,k}\left( 1\right) =l\right) =\mathbf{P}\left( \xi
_{1}=l\right) \frac{\left[ u^{k-l}\right] \Phi \left( u\right) ^{n-1}}{%
\left[ u^{k}\right] \Phi \left( u\right) ^{n}}=
\end{equation*}
\begin{equation*}
\frac{\sigma _{l}\left( \theta \right) x^{l}}{l!}\frac{\left[ u^{k-l}\right]
Z_{\theta }\left( xu\right) ^{n-1}}{\left[ u^{k}\right] Z_{\theta }\left(
xu\right) ^{n}}=\binom{k}{l}\frac{\sigma _{l}\left( \theta \right) \sigma
_{k-l}\left( \left( n-1\right) \theta \right) }{\sigma _{k}\left( n\theta
\right) }.
\end{equation*}
Note that $\sum_{l=0}^{k}\mathbf{P}\left( K_{n,k}\left( 1\right) =l\right)
=1 $, as required, in view of (\ref{f2c}) with $\theta ^{\prime }=\left(
n-1\right) \theta $.\newline

- Proceeding similarly, with $l\in \left\{ 0,...,k\right\} ,$ we would
obtain the law of the partial sums $K_{n,k}\left( 1\right)
+...+K_{n,k}\left( m\right) $, $m<n,$ as 
\begin{equation*}
\mathbf{P}\left( K_{n,k}\left( 1\right) +...+K_{n,k}\left( m\right)
=l\right) =\binom{k}{l}\frac{\sigma _{l}\left( m\theta \right) \sigma
_{k-l}\left( \left( n-m\right) \theta \right) }{\sigma _{k}\left( n\theta
\right) }.
\end{equation*}
As required also, $\sum_{l=0}^{k}\mathbf{P}\left( K_{n,k}\left( 1\right)
+...+K_{n,k}\left( m\right) =l\right) =1$, as a result of $\sigma _{k}\left(
\theta \right) $ being a convolution sequence of polynomials, from (\ref{f2c}%
).\newline

- Finally, define $\left\{ k\right\} _{l}:=k\left( k-1\right) ...\left(
k-l+1\right) $ with $\left\{ k\right\} _{0}:=1$ and let us now consider the
falling factorial moments of $\mathbf{K}_{n,k}$.

Fix $\mathbf{l}_{n}:=\left( l_{1},...,l_{n}\right) \in \Bbb{N}_{0}^{n}$
summing to $l\leq k.$ We have 
\begin{equation*}
\mathbf{E}\left[ \prod_{m=1}^{n}\left\{ K_{n,k}\left( m\right) \right\}
_{l_{m}}\right] =\prod_{m=1}^{n}l_{m}!\frac{\left[ v^{k}\right]
\prod_{m=1}^{n}\left[ v_{m}^{l_{m}}\right] Z_{\theta }\left( xv\left(
v_{m}+1\right) \right) }{\left[ v^{k}\right] Z_{n\theta }\left( xv\right) }.
\end{equation*}
Since $l_{m}!\left[ v_{m}^{l_{m}}\right] Z_{\theta }\left( xv\left(
v_{m}+1\right) \right) =\sum_{k_{m}\geq l_{m}}\frac{\sigma _{k_{m}}\left(
\theta \right) \cdot \left( xv\right) ^{k_{m}}}{\left( k_{m}-l_{m}\right) !}$%
, with $\mathbf{k}_{n}$ summing to $\left| \mathbf{k}_{n}\right| =k$, with $%
\mathbf{k}_{n}\geq \mathbf{l}_{n}$ meaning $k_{1}\geq l_{1},...,k_{n}\geq
l_{n}$, we get 
\begin{equation}
\mathbf{E}\left[ \prod_{m=1}^{n}\left\{ K_{n,k}\left( m\right) \right\}
_{l_{m}}\right] =\frac{\sum_{\mathbf{k}_{n}\geq \mathbf{l}%
_{n}}\prod_{m=1}^{n}\sigma _{k_{m}}\left( \theta \right) /\left(
k_{m}-l_{m}\right) !}{\sigma _{k}\left( n\theta \right) /k!}.  \label{f8}
\end{equation}
These combinatorial quantities arise in the following resampling problem:%
\newline

\textbf{Subsampling without replacement from }$\mathbf{K}_{n,n}.$ Suppose $%
K_{n,n}\left( m\right) $, $m=1,...,n$ are the random box occupancies of some
sample with size exactly equal to the number $n$ of boxes, generated by some
compound-Poisson vector $\mathbf{\xi }_{n}:=\left( \xi _{1},...,\xi
_{n}\right) $. So there are at most $n$ boxes filled by a singleton as a
result of $\sum_{m=1}^{n}K_{n,n}\left( m\right) =n$. Let $p\leq k\leq n.$ In
connection with the theory of compound-Poisson coalescent processes, \cite
{HM}, we are interested in the event that after a random $k-$subsampling
without replacement from $\mathbf{K}_{n,n},$ balls are reassigned at random
into boxes so as to end up in a new occupancy\emph{\ }$\mathbf{K}%
_{n,k}^{\prime }:=\left( K_{n,k}^{\prime }\left( q\right) ;q=1,...,p\right) $%
\emph{\ }where only a fixed number $p$ of the random number $\Pi _{n,k}$ of
filled boxes (labeled in arbitrary order) are occupied. So $\mathbf{K}%
_{n,k}^{\prime }$ obeys\emph{\ }$\sum_{q=1}^{p}K_{n,k}^{\prime }\left(
q\right) =k$ and $K_{n,k}^{\prime }\left( q\right) \geq 1$\emph{. }Then,
with $\left( k_{1},...,k_{p}\right) \in \Bbb{N}^{p}$ summing to $k$, the
sampling without replacement strategy yields: 
\begin{eqnarray*}
\mathbf{P}\left( K_{n,k}^{\prime }\left( 1\right) =k_{1},..,K_{n,k}^{\prime
}\left( p\right) =k_{p};\Pi _{n,k}=p\right) &=&\binom{n}{p}\binom{k}{%
k_{1}..k_{p}}\frac{\mathbf{E}\left( \prod_{q=1}^{p}\left\{ K_{n,n}\left(
q\right) \right\} _{k_{q}}\right) }{\left\{ n\right\} _{k}} \\
&=&\frac{\binom{n}{p}}{\binom{n}{k}}\mathbf{E}\prod_{q=1}^{p}\binom{%
K_{n,n}\left( q\right) }{k_{q}}.
\end{eqnarray*}
Summing over $\left( k_{1},...,k_{p}\right) \in \Bbb{N}^{p}$%
\begin{equation*}
\mathbf{P}\left( \Pi _{n,k}=p\right) =\frac{\binom{n}{p}}{\left\{ n\right\}
_{k}}\sum_{\mathbf{k}_{p}\in \Bbb{N}^{p}:\text{ }\left| \mathbf{k}%
_{p}\right| =k}\binom{k}{k_{1}...k_{p}}\mathbf{E}\left(
\prod_{q=1}^{p}\left\{ K_{n,n}\left( q\right) \right\} _{k_{q}}\right)
\end{equation*}
is the probability that in a $k-$subsampling without replacement from $%
\mathbf{K}_{n,n}$ exactly $p\leq k\leq n$ boxes will be filled. Using (\ref
{f8}), with $\mathbf{k}_{p}:=\left( k_{1},...,k_{p}\right) \in \Bbb{N}^{p}$
a vector of positive integers satisfying $\left| \mathbf{k}_{p}\right|
:=\sum_{q=1}^{p}k_{q}=k$, we have 
\begin{equation*}
\mathbf{E}\left[ \prod_{q=1}^{p}\left\{ K_{n,n}\left( q\right) \right\}
_{k_{q}}\right] =\frac{\sum_{\mathbf{l}_{p}\in \Bbb{N}_{0}^{p}}%
\prod_{q=1}^{p}\sigma _{k_{q}+l_{q}}\left( \theta \right) /l_{q}!}{\sigma
_{n}\left( n\theta \right) /n!},
\end{equation*}
and the full expression of the probabilities $\mathbf{P}\left( \Pi
_{n,k}=p\right) $ can be obtained in terms of the original weights $%
w_{k}\left( \theta \right) =\sigma _{k}\left( \theta \right) /k!.$

These questions arise in the discrete theory of compound-Poisson coalescent
processes. Suppose $\mathbf{K}_{n,n}$ is the random exchangeable
reproduction law of some Markov branching process preserving the total
number $n$ of individuals over the subsequent generations, \cite{HM}. That
is, independently in each generation, individual number $m$ produces $%
K_{n,n}\left( m\right) $ offspring, $m=1,...,n$ and $\sum_{m=1}^{n}K_{n,n}%
\left( m\right) =n.$

We first wish to count, forward in time, the number of descendants of any
size$-m$ subsample of the full population with size $n$, defining thereby a
discrete-time Markov chain. Clearly, the $\left( m,l\right) $ entry of the
transition matrix of this Markov process on the state-space $\left\{
0,...,n\right\} $ is 
\begin{equation*}
\mathbf{P}\left( K_{n,n}\left( 1\right) +...+K_{n,n}\left( m\right)
=l\right) =\binom{n}{l}\frac{\sigma _{l}\left( m\theta \right) \sigma
_{n-l}\left( \left( n-m\right) \theta \right) }{\sigma _{n}\left( n\theta
\right) },\text{ }m,l\in \left\{ 0,...,n\right\} ,
\end{equation*}
looking at the descent of all size$-m$ subsamples. For this Markov chain,
clearly, the states $\left\{ 0,n\right\} $ are both absorbing.

Looking now at this branching process backward in time, individuals are seen
to merge, giving rise to some ancestral coalescent process where individuals
are identified if they share a common ancestor one generation backward in
time. The process stops when a single individual is present (at time to
their most recent common ancestor).

The quantity $\mathbf{P}\left( K_{n,k}^{\prime }\left( 1\right)
=k_{1},...,K_{n,k}^{\prime }\left( p\right) =k_{p};\Pi _{n,k}=p\right) $ is
then the probability that a one-step back $\left( k_{1},...,k_{p}\right) $
to $p$ merger for a subsample of size $k$ occurs in this ancestral process.
The lower-triangular stochastic matrix $\mathcal{Q}_{k,p}^{\left( n\right)
}:=\mathbf{P}\left( \Pi _{n,k}=p\right) $ is the transition matrix of this
pure death coalescent Markov process on $\left\{ 0,...,n\right\} $, with
both states $\left\{ 0,1\right\} $ absorbing. The forward and backward
Markov processes are easily seen to be duals in the sense and for the
duality kernel defined in \cite{Moh}.\emph{\newline
}

\textbf{Number of filled boxes in }$\mathbf{K}_{n,k}$\textbf{:} With $%
\mathbf{I}\left( \cdot \right) $ denoting the indicator function, let now $%
P_{n,k}:=\sum_{m=1}^{n}\mathbf{I}\left( K_{n,k}\left( m\right) >0\right) $
count the number of non empty boxes in the sampling process from $\mathbf{%
\xi }_{n}$. With $1\leq p\leq n\wedge k$, the probability that there are
only $P_{n,k}=p\in \left[ n\right] $ visited boxes in the sampling process,
the $n-p$ remaining ones remaining empty, is easily obtained as follows: In
the event $P_{n,k}=p,$ for any fixed subset $\left( m_{1},...,m_{p}\right) $
of $p$ different box numbers and each $\mathbf{k}_{p}=\left(
k_{1},...,k_{p}\right) $ in $\Bbb{N}^{p}$ summing to $k$, we have from (\ref
{f6}) 
\begin{equation*}
\mathbf{P}\left( \left( K_{n,k}\left( m_{q}\right) =k_{q},\text{ }%
q=1...,p\right) ;P_{n,k}=p\right) =\frac{k!}{\sigma _{k}\left( n\theta
\right) }\prod_{q=1}^{p}\frac{\sigma _{k_{q}}\left( \theta \right) }{k_{q}!}.
\end{equation*}
The above probability is independent of the $\binom{n}{p}$ different subsets 
$\left( m_{1},...,m_{p}\right) $. Denote by $\left\{ L_{1},..,L_{p}\right\} $
the random subset of indexes of the occupied $p$ boxes in the event $%
P_{n,k}=p$. From the above argument, we get 
\begin{equation*}
\mathbf{P}\left( \left\{ K_{n,k}\left( L_{q}\right) ,q=1,...,p\right\}
=\left\{ k_{q},q=1,...,p\right\} ;P_{n,k}=p\right) =\binom{n}{p}\frac{k!}{%
\sigma _{k}\left( n\theta \right) }\prod_{q=1}^{p}\frac{\sigma
_{k_{q}}\left( \theta \right) }{k_{q}!},
\end{equation*}
where $\left\{ K_{n,k}\left( L_{q}\right) ,q=1,...,p\right\} =\left\{
k_{q},q=1,...,p\right\} $ is an equality of multisets (the multisets are
needed to keep in mind the repetitions that could exist in $k_{q},$ $q=1,..,p
$). Letting $\widehat{K}_{n,k}\left( q\right) :=K_{n,k}\left( L_{q}\right) $%
, $q=1,..,p$, the last equality will simply be written as 
\begin{equation}
\mathbf{P}\left( \widehat{K}_{n,k}\left( 1\right) =k_{1},...,\widehat{K}%
_{n,k}\left( p\right) =k_{p};P_{n,k}=p\right) =\binom{n}{p}\frac{k!}{\sigma
_{k}\left( n\theta \right) }\prod_{q=1}^{p}\frac{\sigma _{k_{q}}\left(
\theta \right) }{k_{q}!}.  \label{f9}
\end{equation}
This is the probability that there are $p\in \left[ n\right] $ non-empty
boxes labeled in arbitrary way and that $\left( k_{1},...,k_{p}\right) $ are
their respective occupancies. Note that 
\begin{equation*}
\mathcal{P}_{k,p}^{\left( n\right) }:=\mathbf{P}\left( P_{n,k}=p\right) =%
\binom{n}{p}\frac{k!}{\sigma _{k}\left( n\theta \right) }\sum_{\mathbf{k}%
_{p}\in \Bbb{N}^{p}:\text{ }\left| \mathbf{k}_{p}\right| =k}\prod_{q=1}^{p}%
\frac{\sigma _{k_{q}}\left( \theta \right) }{k_{q}!}
\end{equation*}
is the probability that in a $k-$sample from $n$ species with abundance $%
\mathbf{\xi }_{n}$, the exact number of distinct observed species is $p.$ In
particular, $\mathcal{P}_{k,1}^{\left( n\right) }:=n\frac{\sigma _{k}\left(
\theta \right) }{\sigma _{k}\left( n\theta \right) }$ is the probability
that in this $k-$sample, only one species is discovered (whichever it is).

The lower-triangular stochastic matrix with $\left( k,p\right) $ entries $%
\mathcal{P}_{k,p}^{\left( n\right) }:=\mathbf{P}\left( P_{n,k}=p\right) $ is
the transition matrix of some other pure death Markov process on $\left\{
0,...,n\right\} $ which does not coincide in general with the coalescent
transition matrix $\mathcal{Q}_{k,p}^{\left( n\right) }$ defined in the
latter paragraph (in fact both transition matrices match iff $\xi $ is
negative binomial distributed, see \cite{HM1}).

The expression (\ref{f9}) turns out to be the canonical Gibbs distribution
on finite size-$n$ partitions of $k$ into $p$ distinct clusters (the filled
boxes), derived from the weight sequence $\phi _{\bullet }$. In this
language, the normalizing quantity $\sigma _{k}\left( n\theta \right) /k!$
is called the canonical Gibbs partition function.\newline

Now, from (\ref{f9}), with $\left\{ n\right\} _{p}:=n!/\left( n-p\right) !$%
\begin{equation}
\mathbf{P}\left( P_{n,k}=p\right) =\frac{\left\{ n\right\} _{p}}{\sigma
_{k}\left( n\theta \right) }B_{k,p}\left( \sigma _{\bullet }\left( \theta
\right) \right) ,\text{ }p\in \left\{ 1,...,n\wedge k\right\} ,  \label{f9a}
\end{equation}
where 
\begin{equation}
B_{k,p}\left( \sigma _{\bullet }\left( \theta \right) \right) :=\frac{k!}{p!}%
\sum_{\mathbf{k}_{p}\in \Bbb{N}^{p}:\text{ }\left| \mathbf{k}_{p}\right| =k%
\text{ }}\prod_{q=1}^{p}\frac{\sigma _{k_{q}}\left( \theta \right) }{k_{q}!}=%
\frac{k!}{p!}\left[ x^{k}\right] \left( Z_{\theta }\left( x\right) -1\right)
^{p}  \label{f10}
\end{equation}
is now a Bell polynomial in the polynomial variables $\sigma _{\bullet
}\left( \theta \right) :=\left( \sigma _{1}\left( \theta \right) ,\sigma
_{2}\left( \theta \right) ,...\right) .$

Conditioning the canonical Gibbs distribution on the number of filled cells
being equal to $p$ yields the corresponding micro-canonical distribution as 
\begin{equation*}
\mathbf{P}\left( \widehat{K}_{n,k}\left( 1\right) =k_{1},...,\widehat{K}%
_{n,k}\left( p\right) =k_{p}\mid P_{n,k}=p\right)
\end{equation*}
\begin{equation*}
=\frac{k!}{p!}\frac{1}{B_{k,p}\left( \sigma _{\bullet }\left( \theta \right)
\right) }\prod_{q=1}^{p}\frac{\sigma _{k_{q}}\left( \theta \right) }{k_{q}!}.
\end{equation*}
The new normalizing constant $B_{k,p}\left( \sigma _{\bullet }\left( \theta
\right) \right) /k!$ may be called the microcanonical partition function.

The microcanonical distribution is independent of $n.$ So, for all models
studied here, $P_{n,k}$ is a sufficient statistic in the estimation of $n$
problem from occupancy data (assuming $\theta $ known).\newline

Let us now give some additional details on the distribution of $P_{n,k}$.

\begin{proposition}
$\left( a\right) $ Assume $k\geq n$. The probability generating function of $%
P_{n,k}$ is given by 
\begin{equation}
\mathbf{E}\left( u^{P_{n,k}}\right) =\sum_{p=0}^{n-1}\binom{n}{p}%
u^{n-p}\left( 1-u\right) ^{p}\frac{\sigma _{k}\left( \left( n-p\right)
\theta \right) }{\sigma _{k}\left( n\theta \right) },  \label{f10a}
\end{equation}
with: 
\begin{equation}
\mathbf{P}\left( P_{n,k}=p\right) =\binom{n}{p}\sum_{q=1}^{p}\left(
-1\right) ^{p-q}\binom{p}{q}\frac{\sigma _{k}\left( q\theta \right) }{\sigma
_{k}\left( n\theta \right) },\text{ }p\in \left\{ 1,...,n\right\} .
\label{f10b}
\end{equation}
In addition, 
\begin{equation*}
\mathbf{E}\left( P_{n,k}\right) =n\left( 1-\frac{\sigma _{k}\left( \left(
n-1\right) \theta \right) }{\sigma _{k}\left( n\theta \right) }\right)
\end{equation*}
\begin{equation*}
\text{Var}\left( P_{n,k}\right) =n\left( \frac{\sigma _{k}\left( \left(
n-1\right) \theta \right) }{\sigma _{k}\left( n\theta \right) }+\left(
n-1\right) \frac{\sigma _{k}\left( \left( n-2\right) \theta \right) }{\sigma
_{k}\left( n\theta \right) }-n\left( \frac{\sigma _{k}\left( \left(
n-1\right) \theta \right) }{\sigma _{k}\left( n\theta \right) }\right)
^{2}\right)
\end{equation*}

$\left( b\right) $ If $k<n$, (\ref{f10a}) and (\ref{f10b}) still hold, but
now with a modified support for $P_{n,k}$'s law:\emph{\ } 
\begin{equation}
\mathbf{P}\left( P_{n,k}=p\right) =\binom{n}{p}\sum_{q=1}^{p}\left(
-1\right) ^{p-q}\binom{p}{q}\frac{\sigma _{k}\left( q\theta \right) }{\sigma
_{k}\left( n\theta \right) },\text{ }p\in \left\{ 1,...,k\right\} .
\label{f10b1}
\end{equation}
\end{proposition}

\textbf{Proof}: $\left( a\right) $ This follows from $B_{k,p}\left( \sigma
_{\bullet }\left( \theta \right) \right) =\frac{k!}{p!}\left[ x^{k}\right]
\left( Z_{\theta }\left( x\right) -1\right) ^{p}.$ Indeed, from (\ref{f9a}) 
\begin{equation*}
\mathbf{E}\left( u^{P_{n,k}}\right) =\sum_{p=0}^{n}u^{p}\left\{ n\right\}
_{p}\frac{B_{k,p}\left( \sigma _{\bullet }\left( \theta \right) \right) }{%
\sigma _{k}\left( n\theta \right) }=\frac{k!}{\sigma _{k}\left( n\theta
\right) }\sum_{p=0}^{n}\binom{n}{p}\left[ x^{k}\right] \left( u\left(
Z_{\theta }\left( x\right) -1\right) \right) ^{p}
\end{equation*}
\begin{equation*}
=\frac{k!}{\sigma _{k}\left( n\theta \right) }\left[ x^{k}\right] \left(
1-u+uZ_{\theta }\left( x\right) \right) ^{n}=\frac{k!}{\sigma _{k}\left(
n\theta \right) }\sum_{p=0}^{n}\binom{n}{p}u^{n-p}\left( 1-u\right)
^{p}\left[ x^{k}\right] Z_{\theta }\left( x\right) ^{n-p}
\end{equation*}

\begin{equation*}
=\sum_{p=0}^{n-1}\binom{n}{p}u^{n-p}\left( 1-u\right) ^{p}\frac{\sigma
_{k}\left( \left( n-p\right) \theta \right) }{\sigma _{k}\left( n\theta
\right) }.
\end{equation*}
The alternating sum expression of $\mathbf{P}\left( P_{n,k}=p\right) $
follows from extracting $\left[ u^{p}\right] \mathbf{E}\left(
u^{P_{n,k}}\right) $ and the mean and variance of $P_{n,k}$ from the
evaluations of the first and second derivatives of $\mathbf{E}\left(
u^{P_{n,k}}\right) $ with respect to $u$ at $u=1.$

$\left( b\right) $ follows from similar considerations. Indeed, in
principle, we should start with $\mathbf{E}\left( u^{P_{n,k}}\right)
=\sum_{p=0}^{k}u^{p}\left\{ n\right\} _{p}\frac{B_{k,p}\left( \sigma
_{\bullet }\left( \theta \right) \right) }{\sigma _{k}\left( n\theta \right) 
}$ where the $p-$sum now stops at $p=k=k\wedge n$. But the upper bound of
this $p-$sum can be extended to $n$ because $B_{k,p}\left( \sigma _{\bullet
}\left( \theta \right) \right) =0$ if $p>k$. $\diamond $\newline

In the particular case discussed below when $\sigma _{k}\left( \theta
\right) =\theta \left( \theta +1\right) ...\left( \theta +k-1\right) $
(Ewens-Dirichlet model), these results can be found in \cite{Keener}.

In (\ref{f9a}), the new combinatorial coefficients $B_{k,p}\left( \sigma
_{\bullet }\left( \theta \right) \right) $ come into the game. They are
given by

\begin{corollary}
With $S_{l,p}$ the second kind Stirling numbers, 
\begin{equation*}
B_{k,p}\left( \sigma _{\bullet }\left( \theta \right) \right)
=\sum_{l=p}^{k}B_{k,l}\left( \phi _{\bullet }\right) S_{l,p}\theta
^{l}=\theta ^{p}\cdot \sum_{l=0}^{k-p}B_{k,p+l}\left( \phi _{\bullet
}\right) S_{l+p,p}\theta ^{l},
\end{equation*}
showing that $B_{k,p}\left( \sigma _{\bullet }\left( \theta \right) \right) $
is itself a polynomial in $\theta $ with larger (smaller) degree $k$
(respectively $p$).
\end{corollary}

\textbf{Proof}: From (\ref{f9a}) and (\ref{f10b}), we have (\footnote{%
This identity was derived in a different way in \cite{Yang}.}) 
\begin{equation*}
B_{k,p}\left( \sigma _{\bullet }\left( \theta \right) \right) =\frac{1}{p!}%
\sum_{q=1}^{p}\left( -1\right) ^{p-q}\binom{p}{q}\sigma _{k}\left( q\theta
\right) .
\end{equation*}
Recalling $\sigma _{k}\left( \theta \right) =\sum_{l=1}^{k}\theta
^{l}B_{k,l}\left( \phi _{\bullet }\right) $ and observing $%
S_{l,p}=\sum_{q=1}^{p}\left( -1\right) ^{p-q}\binom{p}{q}q^{l}$ gives the
result after reversing the sums. This result actually is in accordance with
the Faa di Bruno formula (see \cite{Comtet}) giving the Taylor coefficients
of the composition function $g$ of the two analytic functions $g\left(
x\right) :=e_{\lambda ,\theta }\circ \phi \left( x\right) $ where $%
e_{\lambda ,\theta }\left( x\right) :=e^{\lambda \left( e^{\theta
x}-1\right) }$ as 
\begin{equation*}
S_{k}\left( \lambda \right) =\sum_{l=1}^{k}e_{l}\left( \theta ,\lambda
\right) B_{k,l}\left( \phi _{\bullet }\right) ,
\end{equation*}
with $e_{l}\left( \theta ,\lambda \right) =\theta ^{l}\sum_{p=1}^{l}\lambda
^{p}S_{l,p}$ the $l^{\text{th}}$ Taylor coefficient of $e_{\lambda ,\theta
}\left( x\right) .$ Clearly indeed, 
\begin{equation*}
g\left( x\right) =e^{\lambda \left( Z_{\theta }\left( x\right) -1\right)
}=1+\sum_{k\geq 1}\frac{x^{k}}{k!}S_{k}\left( \lambda \right)
=:1+\sum_{k\geq 1}\frac{x^{k}}{k!}\left( \sum_{p=1}^{k}\lambda
^{p}B_{k,p}\left( \sigma _{\bullet }\left( \theta \right) \right) \right)
\end{equation*}
and the $\lambda ^{p}$-coefficient of $S_{k}\left( \lambda \right) $ is
exactly $\sum_{l=p}^{k}B_{k,l}\left( \phi _{\bullet }\right) S_{l,p}\theta
^{l}.$ $\diamond $

\subsection{The estimation of $n$ problem}

Let us now discuss the important question of estimating the unknown number
of species $n$ based on the data $k$ and $P$ (assuming $\theta $ is known),
recalling $\mathbf{P}\left( P_{n,k}=P\right) $ is a sufficient statistic in
this estimation problem$.$ Our forthcoming statement holds for a class of $%
\phi $ which is such that the degree$-k$ polynomial $\sigma _{k}\left(
\theta \right) \in ZR_{-}$(has only real non-positive zeroes). We recall
that $\sigma _{k}\left( \theta \right) \in ZR_{-}$ iff the matrix $M$ with
entries $M_{i,j}=B_{k,i-j}\left( \phi _{\bullet }\right) ,$ $i,j=0,...,k,$
with $B_{k,l}\left( \phi _{\bullet }\right) =0$ if $l\notin \left\{
l:B_{k,l}\left( \phi _{\bullet }\right) >0\right\} $ is totally positive of
order $k$ (with $l=1,...,k,$ each $l\times l$ minor of $M$ has a nonnegative
determinant), \cite{scho}. Therefore, there is no simple way to check
whether or not $\sigma _{k}\left( \theta \right) \in ZR_{-}$.

We also recall here, \cite{scho}, that if and only if the matrix $M=M_{i,j}$
would be such that all its $2\times 2$ minors have a nonnegative
determinant, then the sequence $B_{k,l}\left( \phi _{\bullet }\right) $, $%
l=1,...,k$ (with no internal zeros) is $l-$log-concave (the $l-$sequence $%
B_{k,l}\left( \phi _{\bullet }\right) $ is a P\`{o}lya frequency sequence of
order $2$). If this is the case, we shall say $\sigma _{k}\left( \theta
\right) \in PF_{2}.$

\begin{proposition}
Suppose $\sigma _{k}\left( \theta \right) \in ZR_{-}.$ Then the
log-likelihood $\log \mathbf{P}\left( P_{n,k}=P\right) $ attains its maximum
in $n$ at least once and at most twice in which latter case, the two values
are adjacent integers. This leads to the maximum likelihood estimator $%
\widehat{n}$ of $n$ characterized by: 
\begin{equation*}
\widehat{n}=\sup \left\{ n>0:\frac{\mathbf{P}\left( P_{n,k}=P\right) }{%
\mathbf{P}\left( P_{n-1,k}=P\right) }>1\right\} .
\end{equation*}

When the set of integers $\left\{ n>0:\frac{\mathbf{P}\left(
P_{n,k}=P\right) }{\mathbf{P}\left( P_{n-1,k}=P\right) }>1\right\} $ is
empty, $\widehat{n}=P$.

When this is not the case and for large $n$, an approximation of the
estimator $\widehat{n}$ of $n$ is given by the implicit equation 
\begin{equation*}
P=\widehat{n}\left( 1-\frac{\sigma _{k}\left( \left( \widehat{n}-1\right)
\theta \right) }{\sigma _{k}\left( \widehat{n}\theta \right) }\right) .
\end{equation*}
\end{proposition}

\textbf{Proof}: We extend (\ref{f9a}) to $n$ a real variable, so we can
differentiate $\log \mathbf{P}\left( P_{n,k}=P\right) $ with respect to $%
n>P. $ In this domain, we have $\partial _{n}\log \left\{ n\right\}
_{P}=\sum_{q=0}^{P-1}\frac{1}{n-q}$, and so we get 
\begin{equation*}
\partial _{n}\log \mathbf{P}\left( P_{n,k}=P\right) =\sum_{q=0}^{P-1}\frac{1%
}{n-q}-\partial _{n}\log \sigma _{k}\left( n\theta \right) .
\end{equation*}
Suppose the polynomial $\sigma _{k}\left( \theta \right) \in ZR_{-}$ has
zeroes $-r_{l,k}$ where: $0=r_{1,k}\leq ...\leq r_{k,k}$. Then $\sigma
_{k}\left( n\theta \right) =\prod_{l=1}^{k}\left( n\theta +r_{l,k}\right) $
and $\partial _{n}\log \sigma _{k}\left( n\theta \right)
=\sum_{l=1}^{k}\left( n+r_{l,k}/\theta \right) ^{-1}$, together with $%
\partial _{n}^{2}\log \sigma _{k}\left( n\theta \right)
=-\sum_{l=1}^{k}\left( n+r_{l,k}/\theta \right) ^{-2}<0.$

If $\sum_{q=0}^{P-1}\frac{1}{n-q}-\sum_{l=1}^{k}\left( n+r_{l,k}/\theta
\right) ^{-1}\overset{(*)}{=}0$ , then 
\begin{equation*}
\partial _{n}^{2}\log \mathbf{P}\left( P_{n,k}=P\right) =-\sum_{q=0}^{P-1}%
\frac{1}{\left( n-q\right) ^{2}}+\sum_{l=1}^{k}\left( n+r_{l,k}/\theta
\right) ^{-2}<0,
\end{equation*}
showing that the likelihood is log-concave around the critical points.
Hence, if $\widehat{n}$ solves $(*)$ it is a local maximum and there is no
local minimum. The maximum likelihood estimator of real $n$ is thus unique.

Coming back to $n$ integer, we deduce that the maximum likelihood estimator
of $n$ is the integer $\sup \left\{ n>0:\frac{\mathbf{P}\left(
P_{n,k}=P\right) }{\mathbf{P}\left( P_{n-1,k}=P\right) }>1\right\} $. When $%
n $ is large, it may thus be approximated by $\frac{\mathbf{P}\left( P_{%
\widehat{n},k}=P\right) }{\mathbf{P}\left( P_{\widehat{n}-1,k}=P\right) }=1$%
, leading to 
\begin{equation*}
\frac{\left\{ \widehat{n}\right\} _{P}\sigma _{k}\left( \left( \widehat{n}%
-1\right) \theta \right) }{\left\{ \widehat{n}-1\right\} _{P}\sigma
_{k}\left( \widehat{n}\theta \right) }=1\text{ or }P=\widehat{n}\left( 1-%
\frac{\sigma _{k}\left( \left( \widehat{n}-1\right) \theta \right) }{\sigma
_{k}\left( \widehat{n}\theta \right) }\right) .\text{ }\diamond
\end{equation*}
\newline

\textbf{An alternative estimator.} Let us now come to an alternative
estimator of $n$ (see \cite{Keener} for a similar approach in the particular
context of the Dirichlet model given by $\phi \left( x\right) =-\alpha \log
\left( 1-x\right) $). Suppose that for all $\theta >0$ and $k\geq 1,$ $%
B_{k,p}\left( \sigma _{\bullet }\left( \theta \right) \right) $ is a
log-concave $p-$sequence (equivalently, each degree-$k$ $\lambda -$%
polynomial $S_{k}\left( \lambda \right) \in PF_{2}$)$.$ Then, by Darroch
Theorem \cite{Dar}, $B_{k,p}\left( \sigma _{\bullet }\left( \theta \right)
\right) $ is $p-$unimodal or bimodal at two consecutive $p$. Because the $p-$%
sequence $\left\{ n\right\} _{p}$ is also log-concave, $\left\{ n\right\}
_{p}B_{k,p}\left( \sigma _{\bullet }\left( \theta \right) \right) $ is
itself $p-$log-concave. For each $n$ therefore, there is a unique $%
\widetilde{p}$ defined as $\widetilde{p}=\sup \left\{ p>0:\frac{\mathbf{P}%
\left( P_{n,k}=p\right) }{\mathbf{P}\left( P_{n,k}=p-1\right) }>1\right\} $.
Inverting the map $n\rightarrow \widetilde{p}\left( n\right) $, given $p=P$,
there exists a unique $\widetilde{n},$ approximately characterized by $\frac{%
\mathbf{P}\left( P_{\widetilde{n},k}=P-1\right) }{\mathbf{P}\left( P_{%
\widetilde{n},k}=P\right) }=1$, which can serve as an alternative estimator
of $n$ given the data $\left( k,P\right) .$ From (\ref{f9a}), it is thus
given by 
\begin{equation*}
\widetilde{n}=P+\frac{B_{k,P-1}\left( \sigma _{\bullet }\left( \theta
\right) \right) }{B_{k,P}\left( \sigma _{\bullet }\left( \theta \right)
\right) }.
\end{equation*}

If $k\geq n,$ taking the expectation with respect to $P_{n,k},$ from (\ref
{f10b}), we have 
\begin{eqnarray*}
\mathbf{E}\left( \widetilde{n}\right) &=&\mathbf{E}\left( P_{n,k}\right)
+\sum_{p=1}^{n}\frac{B_{k,p-1}\left( \sigma _{\bullet }\left( \theta \right)
\right) }{B_{k,p}\left( \sigma _{\bullet }\left( \theta \right) \right) }%
\frac{\left\{ n\right\} _{p}}{\sigma _{k}\left( n\theta \right) }%
B_{k,p}\left( \sigma _{\bullet }\left( \theta \right) \right) \\
&=&\mathbf{E}\left( P_{n,k}\right) +\sum_{p=2}^{n}\frac{\left\{ n\right\}
_{p}B_{k,p-1}\left( \phi _{\bullet }\right) }{\sigma _{k}\left( n\theta
\right) }=\mathbf{E}\left( P_{n,k}\right) +\sum_{p=2}^{n}\left( n-\left(
p-1\right) \right) \frac{\left\{ n\right\} _{p-1}B_{k,p-1}\left( \phi
_{\bullet }\right) }{\sigma _{k}\left( n\theta \right) } \\
&=&\mathbf{E}\left( P_{n,k}\right) +n\left( 1-\frac{\left\{ n\right\}
_{n}B_{k,n}\left( \phi _{\bullet }\right) }{\sigma _{k}\left( n\theta
\right) }\right) -\left( \mathbf{E}\left( P_{n,k}\right) -n\frac{\left\{
n\right\} _{n}B_{k,n}\left( \phi _{\bullet }\right) }{\sigma _{k}\left(
n\theta \right) }\right) =n.
\end{eqnarray*}
So, when $k\geq n$, $\widetilde{n}$ is an unbiased estimator of $n$. The
Fisher information of $n$ is 
\begin{equation*}
I\left( n\right) =-\mathbf{E}\left( \partial _{n}^{2}\log \mathbf{P}\left(
P_{n,k}=P\right) \right) =\mathbf{E}\left( \sum_{q=0}^{P-1}\frac{1}{\left(
n-q\right) ^{2}}\right) -\sum_{l=1}^{k}\left( n+r_{l,k}/\theta \right)
^{-2}>0,
\end{equation*}
giving the Cram\'{e}r-Rao bound for the variance: Var$\left( \widetilde{n}%
\right) \geq I\left( n\right) ^{-1}$.

\subsection{Frequency of frequencies}

This suggests to look at the frequency of frequencies distribution problem.
For $i=0,...,k$, let now 
\begin{equation}
A_{n,k}\left( i\right) =\sum_{m=1}^{n}\mathbf{I}\left( K_{n,k}\left(
m\right) =i\right)  \label{f11}
\end{equation}
count the number of boxes visited $i$ times by the $k-$sample, with $%
A_{n,k}\left( 0\right) =n-P_{n,k}$, the number of empty boxes.

Let $\left( a_{1},a_{2},...\right) $ be non-negative integers satisfying $%
\sum_{i\geq 1}a_{i}=p$ and $\sum_{i\geq 1}ia_{i}=k.$

It follows from (\ref{f6}) that 
\begin{equation}
\mathbf{P}\left( A_{n,k}\left( 1\right) =a_{1},A_{n,k}\left( 2\right)
=a_{2},...\right) =\frac{\left\{ n\right\} _{p}\cdot k!}{\sigma _{k}\left(
n\theta \right) }\prod_{i\geq 1}\left\{ \left( \frac{\sigma _{i}\left(
\theta \right) }{i!}\right) ^{a_{i}}\frac{1}{a_{i}!}\right\} .  \label{f11a}
\end{equation}
Taking $A_{n,k}\left( 0\right) $ into account, let $\left(
a_{0},a_{1},...,a_{k}\right) $ be non-negative integers satisfying $%
\sum_{i=0}^{k}a_{i}=n$ and $\sum_{i=1}^{k}ia_{i}=k.$ Then 
\begin{equation*}
\mathbf{P}\left( A_{n,k}\left( 0\right) =a_{0},A_{n,k}\left( 1\right)
=a_{1},...,A_{n,k}\left( k\right) =a_{k}\right) =\frac{n!\cdot k!}{\sigma
_{k}\left( n\theta \right) }\prod_{i=0}^{k}\left\{ \left( \frac{\sigma
_{i}\left( \theta \right) }{i!}\right) ^{a_{i}}\frac{1}{a_{i}!}\right\} .
\end{equation*}
Note from this that, with $\sum_{i=1}^{k}ia_{i}=k$ and $\sum_{1}^{k}a_{i}%
\leq n$, the normalization condition gives the identity 
\begin{equation}
\sum_{a_{1},...,a_{k}}\frac{k!}{\left( n-\sum_{1}^{k}a_{i}\right) !}%
\prod_{i=1}^{k}\left\{ \left( \frac{\sigma _{i}\left( \theta \right) }{i!}%
\right) ^{a_{i}}\frac{1}{a_{i}!}\right\} =\frac{\sigma _{k}\left( n\theta
\right) }{n!}.  \label{ff11}
\end{equation}
From this, we get (see also \cite{PitCSP} Section 1.5):

\begin{proposition}
If $p=n-a_{0}$, the joint distribution of $\left( A_{n,k}\left( 1\right)
,...,A_{n,k}\left( k\right) \right) $ and $P_{n,k}$ reads 
\begin{equation}
\mathbf{P}\left( A_{n,k}\left( 1\right) =a_{1},...,A_{n,k}\left( k\right)
=a_{k};P_{n,k}=p\right) =\frac{\left\{ n\right\} _{p}\cdot k!}{\sigma
_{k}\left( n\theta \right) }\prod_{i=1}^{k}\left\{ \left( \frac{\sigma
_{i}\left( \theta \right) }{i!}\right) ^{a_{i}}\frac{1}{a_{i}!}\right\} .
\label{f12}
\end{equation}
\end{proposition}

Let us compute the falling factorial moments of $A_{n,k}\left( i\right) $, $%
i=1,...,k.$

\begin{proposition}
Let $r_{i},$ $i=1,...,k$ be non-negative integers satisfying $%
\sum_{1}^{k}r_{i}=r\leq n$ and $\sum_{1}^{k}ir_{i}=\kappa \leq k$. We have 
\begin{equation}
\mathbf{E}\left[ \prod_{i=1}^{k}\left\{ A_{n,k}\left( i\right) \right\}
_{r_{i}}\right] =\left\{ n\right\} _{r}\left\{ k\right\} _{\kappa }\frac{%
\sigma _{k-\kappa }\left( \left( n-r\right) \theta \right) }{\sigma
_{k}\left( n\theta \right) }\prod_{i=1}^{k}\left( \frac{\sigma _{i}\left(
\theta \right) }{i!}\right) ^{r_{i}}.  \label{f13}
\end{equation}
\end{proposition}

\textbf{Proof: } 
\begin{equation*}
\mathbf{E}\left[ \prod_{i=1}^{k}\left\{ A_{n,k}\left( i\right) \right\}
_{r_{i}}\right] =\frac{n!\cdot k!}{\sigma _{k}\left( n\theta \right) }%
\sum_{a_{1},...,a_{k}}\frac{1}{\left( n-\sum_{1}^{k}a_{i}\right) !}%
\prod_{i=1}^{k}\left\{ \left( \frac{\sigma _{i}\left( \theta \right) }{i!}%
\right) ^{a_{i}}\frac{1}{\left( a_{i}-r_{i}\right) !}\right\}
\end{equation*}
\begin{equation*}
=\frac{n!\cdot k!}{\sigma _{k}\left( n\theta \right) }\prod_{i=1}^{k}\left( 
\frac{\sigma _{i}\left( \theta \right) }{i!}\right)
^{r_{i}}\sum_{a_{1},...,a_{k}}\frac{1}{\left( n-\sum_{1}^{k}a_{i}\right) !}%
\prod_{i=1}^{k}\left\{ \left( \frac{\sigma _{i}\left( \theta \right) }{i!}%
\right) ^{a_{i}-r_{i}}\frac{1}{\left( a_{i}-r_{i}\right) !}\right\} .
\end{equation*}
The normalization condition (\ref{ff11}) gives: 
\begin{equation*}
\sum_{a_{1},...,a_{k}}\frac{1}{\left( n-\sum_{1}^{k}a_{i}\right) !}%
\prod_{i=1}^{k}\left\{ \left( \frac{\sigma _{i}\left( \theta \right) }{i!}%
\right) ^{a_{i}-r_{i}}\frac{1}{\left( a_{i}-r_{i}\right) !}\right\} =\frac{%
\sigma _{k-\kappa }\left( \left( n-r\right) \theta \right) }{\left(
n-r\right) !\cdot \left( k-\kappa \right) !}.
\end{equation*}
Finally, we get 
\begin{equation*}
\mathbf{E}\left[ \prod_{i=1}^{k}\left\{ A_{n,k}\left( i\right) \right\}
_{r_{i}}\right] =\left\{ n\right\} _{r}\left\{ k\right\} _{\kappa }\frac{%
\sigma _{k-\kappa }\left( \left( n-r\right) \theta \right) }{\sigma
_{k}\left( n\theta \right) }\prod_{i=1}^{k}\left( \frac{\sigma _{i}\left(
\theta \right) }{i!}\right) ^{r_{i}}.\text{ }\diamond
\end{equation*}
\newline

In particular, if all $r_{i}=0$, except for one $i$ for which $r_{i}=1$ ($%
r=1,$ $\kappa =i$), then 
\begin{equation}
\mathbf{E}\left[ A_{n,k}\left( i\right) \right] =n\left\{ k\right\} _{i}%
\frac{\sigma _{k-i}\left( \left( n-1\right) \theta \right) }{\sigma
_{k}\left( n\theta \right) }\frac{\sigma _{i}\left( \theta \right) }{i!}=n%
\mathbf{P}\left( K_{n,k}\left( 1\right) =i\right) .  \label{f14}
\end{equation}
This shows that the expected number of cells visited $i$ times is $n$ times
the probability that there are $i$ visits to (say) cell one. In fact, we
have the more general statement (see also \cite{PitCSP} Section 1.5):

\begin{corollary}
If $r_{i}=\#\left\{ m\in \left\{ 1,...,n\right\} :k_{m}=i\right\} ,$ then 
\begin{equation*}
\mathbf{E}\left[ \prod_{i=1}^{k}\left\{ A_{n,k}\left( i\right) \right\}
_{r_{i}}\right] =n!\mathbf{P}\left( K_{n,k}\left( 1\right)
=k_{1},...,K_{n,k}\left( n\right) =k_{n}\right) ,
\end{equation*}
so that the joint falling factorial moments of the $A$'s can directly be
obtained in terms of the joint distribution of the $K$'s$.$
\end{corollary}

\textbf{Proof:} With the $r_{i}$ as stated, using a sampling without
replacement argument 
\begin{equation*}
\mathbf{P}\left( K_{n,k}\left( 1\right) =k_{1},...,K_{n,k}\left( n\right)
=k_{n}\mid A_{n,k}\left( 1\right) ,...,A_{n,k}\left( k\right) \right) =
\end{equation*}
\begin{equation*}
\frac{1}{n!}\prod_{i=1}^{k}\left\{ A_{n,k}\left( i\right) \right\} _{r_{i}}.
\end{equation*}
Averaging over the $A$'s gives the announced result. $\diamond $

\subsection{The $*-$limit of sampling distributions (the infinitely many
species abundance model)}

Theoretical biologists work in a framework of a population with infinitely
many species, with the more frequent one occurring with abundance $\xi
_{\left( 1\right) },$ second more frequent with abundance $\xi _{\left(
2\right) },...$ with $\xi _{\left( 1\right) }\geq \xi _{\left( 2\right)
}\geq $... Sampling from $\left( \xi _{\left( 1\right) },\xi _{\left(
2\right) },...\right) $ turns out to be a challenging problem. This requires
the introduction of a model with infinitely many species (not only $n$) with
ordered abundance $\xi _{\left( m\right) }$, $m\geq 1.$ For such abundance
models, a $k-$sample will represent the met individuals of various species
when sampling from a population with infinitely many species, \cite{Bunge}.
One can think of obtaining such models while considering the limit $%
n\rightarrow \infty $ and $\theta \rightarrow 0$ in the finite model with $n$
species. Indeed, as we saw, small values of the temperature $\theta >0$ was
an indication on how disparate the abundance numbers $\mathbf{\xi }_{n}$
were. Then, although (as a result of $\mathbf{P}\left( \xi _{1}=0\right)
=\sigma _{0}\left( \theta \right) /Z_{\theta }\left( x\right) \underset{%
\theta \rightarrow 0}{\rightarrow }1$) the $\left( \xi _{m}\right)
_{m=1}^{n} $ are all small in the limit, there is some hope that sampling
from the ranked $\xi _{\left( m\right) }$'s would have a non-degenerate
limit as $n\rightarrow \infty $, $\theta \rightarrow 0$ while $n\theta
\rightarrow \gamma >0.$ We call such a limit the $*-$limit.\newline

It turns out that for the class of Gibbs-Poisson allocation models
considered in this Section, the $*-$limit always makes sense. This
illustrates that limiting models should come down from some finitary
counterpart, \cite{GS}. We first verify our claim intuitively (see also \cite
{PitCSP}, Section $1.5$). Observing indeed that 
\begin{equation*}
\sigma _{k}\left( \theta \right) \sim _{\theta \downarrow 0}\theta
B_{k,1}\left( \phi _{\bullet }\right) =\theta \phi _{k}\text{ and }%
B_{k,p}\left( \sigma _{\bullet }\left( \theta \right) \right) \sim _{\theta
\downarrow 0}\theta ^{p}B_{k,p}\left( \phi _{\bullet }\right)
\end{equation*}
and recalling $\left\{ n\right\} _{p}\sim _{n\rightarrow \infty }n^{p}$, we
easily get:

\begin{proposition}
From (\ref{f9}), with $\left( k_{1},...,k_{p}\right) \in \Bbb{N}^{p}$
summing to $k$ and $p\leq k$%
\begin{equation*}
\mathbf{P}\left( \widehat{K}_{n,k}\left( 1\right) =k_{1},...,\widehat{K}%
_{n,k}\left( p\right) =k_{p};P_{n,k}=p\right) \rightarrow _{*}
\end{equation*}
\begin{equation}
\mathbf{P}^{*}\left( \widehat{K}_{k}\left( 1\right) =k_{1},...,\widehat{K}%
_{k}\left( p\right) =k_{p};P_{k}=p\right) =\frac{k!}{p!}\frac{\gamma ^{p}}{%
\sigma _{k}\left( \gamma \right) }\prod_{q=1}^{p}\frac{\phi _{k_{q}}}{k_{q}!}
\label{f15}
\end{equation}
and, from (\ref{f9a}, \ref{f10}) 
\begin{equation}
\mathbf{P}\left( P_{n,k}=p\right) \rightarrow _{*}\mathbf{P}^{*}\left(
P_{k}=p\right) =\frac{\gamma ^{p}}{\sigma _{k}\left( \gamma \right) }%
B_{k,p}\left( \phi _{\bullet }\right) .  \label{f16}
\end{equation}
Equivalently, the limiting probability generating function of $P_{k}$ also
reads 
\begin{equation}
\mathbf{E}^{*}\left( u^{P_{k}}\right) =\frac{\sigma _{k}\left( \gamma
u\right) }{\sigma _{k}\left( \gamma \right) },  \label{f17}
\end{equation}
with mean $\mathbf{E}^{*}\left( P_{k}\right) =\gamma \frac{\sigma
_{k}^{\prime }\left( \gamma \right) }{\sigma _{k}\left( \gamma \right) }$.
From this, 
\begin{equation}
\mathbf{P}^{*}\left( \widehat{K}_{k}\left( 1\right) =k_{1},...,\widehat{K}%
_{k}\left( p\right) =k_{p}\mid P_{k}=p\right) =\frac{k!}{p!}\frac{1}{%
B_{k,p}\left( \phi _{\bullet }\right) }\prod_{q=1}^{p}\frac{\phi _{k_{q}}}{%
k_{q}!}  \label{f15a}
\end{equation}
which is independent of $\gamma .$

Further, from (\ref{f11a}), with $\left( a_{1},a_{2},...\right) $ satisfying 
$\sum_{i\geq 1}ia_{i}=k$ and $\sum_{i\geq 1}a_{i}=p$%
\begin{equation*}
\mathbf{P}\left( A_{n,k}\left( 1\right) =a_{1},A_{n,k}\left( 2\right)
=a_{2},...\right) \rightarrow _{*}
\end{equation*}
\begin{equation}
\mathbf{P}^{*}\left( A_{k}\left( 1\right) =a_{1},A_{k}\left( 2\right)
=a_{2},...\right) =\frac{\gamma ^{p}k!}{\sigma _{k}\left( \gamma \right) }%
\prod_{i=1}^{k}\frac{\left( \phi _{i}/i!\right) ^{a_{i}}}{a_{i}!}.
\label{f15b}
\end{equation}

Equivalently, from (\ref{f12}) 
\begin{equation*}
\mathbf{P}\left( A_{n,k}\left( 1\right) =a_{1},...,A_{n,k}\left( k\right)
=a_{k};P_{n,k}=p\right) \rightarrow _{*}
\end{equation*}
\begin{equation}
\mathbf{P}^{*}\left( A_{k}\left( 1\right) =a_{1},...,A_{k}\left( k\right)
=a_{k};P_{k}=p\right) =\frac{\gamma ^{p}k!}{\sigma _{k}\left( \gamma \right) 
}\prod_{i=1}^{k}\frac{\left( \phi _{i}/i!\right) ^{a_{i}}}{a_{i}!}.
\label{f19}
\end{equation}
and 
\begin{equation}
\mathbf{P}^{*}\left( A_{k}\left( 1\right) =a_{1},...,A_{k}\left( k\right)
=a_{k}\mid P_{k}=p\right) =\frac{k!}{B_{k,p}\left( \phi _{\bullet }\right) }%
\prod_{i=1}^{k}\frac{\left( \phi _{i}/i!\right) ^{a_{i}}}{a_{i}!},
\label{f19a}
\end{equation}
which is also independent of $\gamma $.
\end{proposition}

(\ref{f15}) or (\ref{f19}) are the canonical Gibbs distributions on
partitions of $k$ into $p$ distinct clusters, derived from the weight
sequence $\phi _{\bullet }$. In this context, the normalizing quantity $%
\sigma _{k}\left( \gamma \right) /k!$ is called the canonical Gibbs
partition polynomial(\footnote{%
The occupancy distribution (\ref{f19}) also appears in Ecology in a species
abundance model occurring in the Hubbell's unified neutral theory of
biodiversity. In this context, $\gamma $ is the fundamental biodiversity
number, \cite{Hub}.}). Conditioning the canonical Gibbs distribution on the
number of filled boxes being equal to $p$ yields the corresponding
micro-canonical distributions (\ref{f15a}) or (\ref{f19a}). The new
normalizing constant $B_{k,p}\left( \phi _{\bullet }\right) /k!$ is called
the microcanonical partition function.

Let us finally compute the falling factorial moments of $A_{k}\left(
i\right) $, $i=1,...,k.$

\begin{proposition}
Let $r_{i},$ $i=1,...,k$ be non-negative integers satisfying $%
\sum_{1}^{k}r_{i}=r$ and $\sum_{1}^{k}ir_{i}=\kappa \leq k$. We have 
\begin{equation}
\mathbf{E}^{*}\left[ \prod_{i=1}^{k}\left\{ A_{k}\left( i\right) \right\}
_{r_{i}}\right] =\gamma ^{r}\left\{ k\right\} _{\kappa }\frac{\sigma
_{k-\kappa }\left( \gamma \right) }{\sigma _{k}\left( \gamma \right) }%
\prod_{i=1}^{k}\left( \frac{\phi _{i}}{i!}\right) ^{r_{i}}.  \label{f19b}
\end{equation}
\end{proposition}

\textbf{Proof: }This follows straightforwardly from\textbf{\ }Proposition $5$
while taking the $*-$limit and using $\sigma _{i}\left( \theta \right) \sim
\theta \phi _{i}$ for small $\theta .$ This formula is a generalization of
the Watterson expression \cite{Wat} obtained in the special Ewens context
when $\phi \left( x\right) =-\log \left( 1-x\right) $, with $\phi
_{i}=\left( i-1\right) !$ and $\sigma _{k}\left( \gamma \right) =\Gamma
\left( \gamma +k\right) /\Gamma \left( \gamma \right) =:\left( \gamma
\right) _{k};$ see Section $3$ for a special account on this model. From (%
\ref{f19b}), we easily get a closed-form expression for the mean $\mathbf{E}%
^{*}\left( A_{k}\left( i\right) \right) $, $i\leq k$, the variance Var$%
^{*2}\left( A_{k}\left( i\right) \right) $, for all $i$ with $2i\leq k$ and
the covariance Cov$^{_{*}}\left( A_{k}\left( i_{1}\right) ,A_{k}\left(
i_{2}\right) \right) $ for all $i_{1}\neq i_{2}$, $i_{1}+i_{2}\leq k.$ $%
\diamond $\newline

We observed that (\ref{f15a}) or (\ref{f19a}) were independent of $\gamma ,$
meaning that $P_{k}$ is a sufficient statistic in the estimation of $\gamma $
problem. Let us now briefly investigate this problem.

\subsection{The estimation of $\gamma $ problem.}

We wish now to discuss the question of estimating $\gamma $ from the data $k$
and $P.$ From (\ref{f16}) 
\begin{equation*}
\partial _{\gamma }\log \mathbf{P}^{*}\left( P_{k}=P\right) =P/\gamma
-\partial _{\gamma }\log \sigma _{k}\left( \gamma \right) .
\end{equation*}
Suppose the polynomial $\sigma _{k}\left( \gamma \right) \in ZR_{-}$ with
zeroes $-r_{l,k}$ where: $0=r_{1,k}\leq ...\leq r_{k,k}$. Then $\sigma
_{k}\left( \gamma \right) =\prod_{l=1}^{k}\left( \gamma +r_{l,k}\right) $
and $\partial _{\gamma }\log \sigma _{k}\left( \gamma \right)
=\sum_{l=1}^{k}\left( \gamma +r_{l,k}\right) ^{-1}$, together with $\partial
_{\gamma }^{2}\log \sigma _{k}\left( \gamma \right) =-\sum_{l=1}^{k}\left(
\gamma +r_{l,k}\right) ^{-2}<0$ ($\gamma \rightarrow \sigma _{k}\left(
\gamma \right) $ is log-concave)$.$

If $P/\gamma -\sum_{l=1}^{k}\left( \gamma +r_{l,k}\right) ^{-1}\overset{(*)}{%
=}0$ , then 
\begin{equation*}
\partial _{\gamma }^{2}\log \mathbf{P}^{*}\left( P_{k}=P\right) =-P/\gamma
^{2}+\sum_{l=1}^{k}\left( \gamma +r_{l,k}\right) ^{-2}<0,
\end{equation*}
showing that $\widehat{\gamma }$ solving $(*)$ is a local maximum and that $%
\log \mathbf{P}^{*}\left( P_{k}=P\right) $ has no local minima. So $\widehat{%
\gamma }$ is the maximum likelihood estimator of $\gamma .$ Even though $%
\sigma _{k}\left( \gamma \right) $ ($1/\sigma _{k}\left( \gamma \right) $)
is a log-concave (respectively log-convex) function of $\gamma $, the
log-likelihood is a log-concave function of $\gamma $ leading to the
existence of $\widehat{\gamma }.$ To summarize, there exists a maximum
likelihood estimator $\widehat{\gamma }$ of $\gamma $ which is characterized
by the implicit equation: 
\begin{equation*}
P=\widehat{\gamma }\frac{\sigma _{k}^{\prime }\left( \widehat{\gamma }%
\right) }{\sigma _{k}\left( \widehat{\gamma }\right) }.
\end{equation*}
\newline

Let us now come to another estimator of $\gamma $. If $\sigma _{k}\left(
\gamma \right) \in ZR_{-},$ then by Newton's inequality (\cite{HLP}, p.52) 
\begin{equation*}
B_{k,p}\left( \phi _{\bullet }\right) ^{2}\geq B_{k,p-1}\left( \phi
_{\bullet }\right) B_{k,p+1}\left( \phi _{\bullet }\right) \left( 1+\frac{1}{%
p}\right) \left( 1+\frac{1}{k-p}\right) >B_{k,p-1}\left( \phi _{\bullet
}\right) B_{k,p+1}\left( \phi _{\bullet }\right) .
\end{equation*}
So $B_{k,p}\left( \phi _{\bullet }\right) $ is $p-$log-concave and by
Darroch Theorem, $B_{k,p}\left( \phi _{\bullet }\right) $ is $p-$unimodal or
bimodal at two consecutive $p$, with mode (maybe up to one unit) equal to $%
\sigma _{k}^{\prime }\left( 1\right) /\sigma _{k}\left( 1\right) $. Because
the $p-$sequence $\gamma ^{p}$ is also log-concave (and log-convex), $\gamma
^{p}B_{k,p}\left( \phi _{\bullet }\right) $ is itself $p-$log-concave and
therefore there exists a unique $\widetilde{\gamma }$ such that $\frac{%
\mathbf{P}^{*}\left( P_{k}=P\right) }{\mathbf{P}^{*}\left( P_{k}=P-1\right) }%
=1.$ It is thus defined by 
\begin{equation*}
\frac{\gamma ^{p}B_{k,P}\left( \phi _{\bullet }\right) }{\gamma
^{p-1}B_{k,P-1}\left( \phi _{\bullet }\right) }=1,\text{ or }\widetilde{%
\gamma }=\frac{B_{k,P-1}\left( \phi _{\bullet }\right) }{B_{k,P}\left( \phi
_{\bullet }\right) }.
\end{equation*}
This $\widetilde{\gamma }$ is an alternative explicit estimator of $\gamma $
based on the data $k$ and $P.$

Taking the expectation with respect to $P_{k},$ we have 
\begin{eqnarray*}
\mathbf{E}^{*}\left( \widetilde{\gamma }\right) &=&\sum_{p=1}^{k}\frac{%
B_{k,p-1}\left( \phi _{\bullet }\right) }{B_{k,p}\left( \phi _{\bullet
}\right) }\frac{\gamma ^{p}}{\sigma _{k}\left( \gamma \right) }B_{k,p}\left(
\phi _{\bullet }\right) =\gamma \sum_{p=2}^{k}B_{k,p-1}\left( \phi _{\bullet
}\right) \frac{\gamma ^{p-1}}{\sigma _{k}\left( \gamma \right) } \\
&=&\gamma \sum_{p=1}^{k-1}B_{k,p}\left( \phi _{\bullet }\right) \frac{\gamma
^{p}}{\sigma _{k}\left( \gamma \right) }=\gamma \left( 1-\frac{\left( \phi
_{1}\gamma \right) ^{k}}{\sigma _{k}\left( \gamma \right) }\right) <\gamma .
\end{eqnarray*}
This shows that $\widetilde{\gamma }$ is not an unbiased estimator of $%
\gamma $.

\textbf{Remark:} The estimator $\widetilde{\gamma }$ only requires that the
sequence $B_{k,p}\left( \phi _{\bullet }\right) $ be $p-$log-concave and,
although sufficient, it is therefore not necessary that $\sigma _{k}\left(
\gamma \right) \in ZR_{-};$ the sequence $B_{k,p}\left( \phi _{\bullet
}\right) $ only needs to be a P\`{o}lya frequency sequence of order $2$ (so $%
\sigma _{k}\left( \gamma \right) \in PF_{2}$) for $\widetilde{\gamma }$ to
be well-defined. In this spirit, we draw the attention on a result in \cite
{BDN}, stating that if the non-null roots of $\sigma _{k}\left( \gamma
\right) $ all lie in the angular cone $\phi \in \left( 2\pi /3,4\pi
/3\right) $ of the complex plane, then $\sigma _{k}\left( \gamma \right) $
has $p-$log-concave coefficients. See \cite{IP} for a bulk of work
pertaining to the diversity estimation parameter for Gibbs partitions.

\section{Sampling from Dirichlet partition: a special case}

We now briefly investigate one particular model of species abundance $%
\mathbf{\xi }_{n}$.\newline

$\bullet $ \textbf{Sampling from a negative binomial sample.}

Assume $\phi \left( x\right) =-\log \left( 1-x\right) ,$ with $\phi
_{m}=\left( m-1\right) !$ and let $Z_{\theta }\left( x\right) =\left(
1-x\right) ^{-\theta }$. Thus, with $\left( \theta \right) _{k}:=\theta
\left( \theta +1\right) ...\left( \theta +k-1\right) $ denoting the (rising
factorial) Pochhammer symbol, $\sigma _{k}\left( \theta \right) =\left(
\theta \right) _{k}$ and $\xi $ is a negative binomial random variable with
parameters $\theta $ and $1-x$. Note that $\sigma _{k}\left( \theta \right)
\in ZR_{-}.$ From (\ref{f5}), the jumps' height $\delta $ of $\xi $ is seen
to obey a logarithmic series distribution.

When sampling from this discrete species-abundance model $\mathbf{\xi }%
_{n}=\left( \xi _{1},...,\xi _{n}\right) $, for instance (\ref{f6}) takes
the particular form:

\begin{equation}
\mathbf{P}\left( \mathbf{K}_{n,k}=\mathbf{k}_{n}\right) =\frac{\mathbf{P}%
\left( \xi _{1}=k_{1},...,\xi _{n}=k_{n}\right) }{\mathbf{P}\left( \zeta
_{n}=k\right) }=\frac{k!}{\left( n\theta \right) _{k}}\prod_{m=1}^{n}\frac{%
\left( \theta \right) _{k_{m}}}{k_{m}!}.  \label{f6d}
\end{equation}
Substituting $\left( \theta \right) _{k}$ to $\sigma _{k}\left( \theta
\right) $ in (\ref{f9}) gives its particular expression.

Because $\sigma _{k+1}\left( \theta \right) =\left( k+\theta \right) \sigma
_{k}\left( \theta \right) ,$ it follows from (\ref{f2}) and (\ref{f2b}) that
with $S_{k}\left( \lambda \right) =k!\left[ x^{k}\right] e^{\lambda \left(
\left( 1-x\right) ^{-\theta }-1\right) }$, $S_{k+1}\left( \lambda \right)
=\left( \theta \lambda +k\right) S_{k}\left( \lambda \right) +\theta \lambda
S_{k}^{\prime }\left( \lambda \right) .$ Thus, the Bell coefficients $%
B_{k,p}\left( \sigma _{\bullet }\left( \theta \right) \right) =B_{k,p}\left(
\left( \theta \right) _{\bullet }\right) =\left[ \lambda ^{p}\right]
S_{k}\left( \lambda \right) ,$ appearing in (\ref{f9a}), obey a simple $3-$%
term recurrence \cite{Donelly}, \cite{Hui} 
\begin{equation*}
B_{k+1,p}\left( \left( \theta \right) _{\bullet }\right) =\theta
B_{k,p-1}\left( \left( \theta \right) _{\bullet }\right) +\left( p\theta
+k\right) B_{k,p}\left( \left( \theta \right) _{\bullet }\right) \text{,}
\end{equation*}
which should be considered with the boundary conditions 
\begin{equation*}
B_{k,0}\left( \left( \theta \right) _{\bullet }\right) =B_{0,p}\left( \left(
\theta \right) _{\bullet }\right) =0,
\end{equation*}
except for $B_{0,0}\left( \left( \theta \right) _{\bullet }\right) :=1.$
This observation is important because it follows from (\ref{f9a}), that,
there exist transition probabilities 
\begin{equation*}
\mathbf{P}\left( P_{n,k+1}=p+1\mid P_{n,k}=p\right) =\frac{\left( n-p\right)
\theta }{n\theta +k}\text{ and}
\end{equation*}
\begin{equation*}
\mathbf{P}\left( P_{n,k+1}=p\mid P_{n,k}=p\right) =\frac{\sum_{r=1}^{p}%
\left( \theta +k_{r}\right) }{n\theta +k}=\frac{p\theta +k}{n\theta +k}.
\end{equation*}
such that, 
\begin{equation*}
\mathbf{P}\left( P_{n,k+1}=p\right) =\frac{\left( n-p+1\right) \theta }{%
n\theta +k}\mathbf{P}\left( P_{n,k}=p-1\right) +\frac{p\theta +k}{n\theta +k}%
\mathbf{P}\left( P_{n,k}=p\right) .
\end{equation*}
The first transition probability gives the probability of the event that a
new species is discovered given $p<n$ of them were discovered from a
previous sample of size $k\geq p$ (the so-called law of succession, \cite{Ew}%
) in a population with $n$ species. Note that $P_{n,k}$ is a Markov chain in 
$k.$

Considering the sampling formulae in the $*-$limit, the expressions (\ref
{f15a}) and (\ref{f19a}) with $\phi _{i}=\left( i-1\right) !$ and $%
B_{k,p}\left( \phi _{\bullet }\right) =s_{k,p}$ (the absolute first kind
Stirling numbers) are the Ewens sampling formulae \cite{Ewen}. Due to $%
\sigma _{k+1}\left( \theta \right) =\left( k+\theta \right) \sigma
_{k}\left( \theta \right) ,$ the Bell coefficients $B_{k,p}\left( \phi
_{\bullet }\right) =B_{k,p}\left( \left( \bullet -1\right) !\right) $ also
obey a $3-$term recurrence 
\begin{equation*}
B_{k+1,p}\left( \left( \bullet -1\right) !\right) =B_{k,p-1}\left( \left(
\bullet -1\right) !\right) +kB_{k,p}\left( \left( \bullet -1\right) !\right) 
\text{.}
\end{equation*}
\newline

$\bullet $ \textbf{Sampling from a symmetric Dirichlet prior.}

It turns out that this sampling formula can be obtained while following a
different path for the sampling procedure:

Consider indeed the following random partition into $n$ fragments of the
unit interval. Let $\theta >0$ be some parameter and assume that the random
fragments sizes $\mathbf{S}_{n}\left( \theta \right) :=\left( S_{1,\theta
},...,S_{n,\theta }\right) $ (with $\sum_{m=1}^{n}S_{m,\theta }=1$) are
distributed according to the (exchangeable) Dirichlet $D_{n}\left( \theta
\right) $ density function on the $n-$simplex, that is to say 
\begin{equation}
f_{S_{1,\theta },...,S_{n,\theta }}\left( s_{1},...,s_{n}\right) =\frac{%
\Gamma \left( n\theta \right) }{\Gamma \left( \theta \right) ^{n}}%
\prod_{m=1}^{n}s_{m}^{\theta -1}\cdot \delta _{\left(
\sum_{m=1}^{n}s_{m}=1\right) }.  \label{eq1}
\end{equation}
Alternatively, with $\left( \theta \right) _{q}:=\Gamma \left( \theta
+q\right) /\Gamma \left( \theta \right) ,$ the law of $\mathbf{S}_{n}\left(
\theta \right) $ is characterized by its joint moment function 
\begin{equation}
\mathbf{E}\left( \prod_{m=1}^{n}S_{m,\theta }^{q_{m}}\right) =\frac{1}{%
\left( n\theta \right) _{\sum_{m=1}^{n}q_{m}}}\prod_{m=1}^{n}\left( \theta
\right) _{q_{m}}.  \label{eq1a}
\end{equation}
We shall put $\mathbf{S}_{n}\left( \theta \right) \overset{d}{\sim }$ $%
D_{n}\left( \theta \right) $ if $\mathbf{S}_{n}\left( \theta \right) $ is
Dirichlet distributed with parameter $\theta $. $\mathbf{S}_{n}\left( \theta
\right) $ can be obtained while considering $\left( Y_{\theta }\overset{d}{=}%
Y_{1,\theta },...,Y_{n,\theta }\right) $, an iid random vector with $%
Y_{\theta }\overset{d}{\sim }$ gamma$\left( \theta \right) $ and letting $%
S_{m,\theta }=Y_{m,\theta }/\left( Y_{1,\theta }+...+Y_{n,\theta }\right) $, 
$m=1,...,n$ (normalizing the $Y_{m,\theta }$'s by their sum)$.$ $\mathbf{S}%
_{n}\left( \theta \right) $ accounts now for a $n-$species frequency
(proportion) model, but now in the continuum. We now come to the sampling
procedure from $\mathbf{S}_{n}\left( \theta \right) $.

Let $\left( U_{1},...,U_{k}\right) $ be $k$ iid uniform throws on the unit
interval partitioned according to $\mathbf{S}_{n}\left( \theta \right) $.
Let 
\begin{equation*}
\mathbf{K}_{n,k}:=\left( K_{n,k}\left( 1\right) ,...,K_{n,k}\left( n\right)
\right)
\end{equation*}
be an integral-valued random vector which counts the number of visits to the
different fragments of $\mathbf{S}_{n}\left( \theta \right) $ in this $k-$%
sample. Hence, if $M_{l}$ is the random fragment label in which the $l^{%
\text{th}}$ trial $U_{l}$ falls, $K_{n,k}\left( m\right) :=\sum_{l=1}^{k}%
\mathbf{I}\left( M_{l}=m\right) $, $m=1,...,n.$

With $\left| \mathbf{k}_{n}\right| =k$ and $\mathbf{k}_{n}:=\left(
k_{1},...,k_{n}\right) \in \Bbb{N}_{0}^{n}$ the non-negative occupancy
vector, as sampling in terms of uniforms $U_{l}$ is equivalent to the
multinomial, $\mathbf{K}_{n,k}$ follows the conditional multinomial
distribution: 
\begin{equation}
\mathbf{P}\left( \mathbf{K}_{n,k}=\mathbf{k}_{n}\mid \mathbf{S}_{n}\left(
\theta \right) \right) =\frac{k!}{\prod_{m=1}^{n}k_{m}!}\prod_{m=1}^{n}S_{m,%
\theta }^{k_{m}}.  \label{e0}
\end{equation}
Averaging over $\mathbf{S}_{n}\left( \theta \right) $, we find 
\begin{equation}
\mathbf{P}\left( \mathbf{K}_{n,k}=\mathbf{k}_{n}\right) =\mathbf{EP}\left( 
\mathbf{K}_{n,k}=\mathbf{k}_{n}\mid \mathbf{S}_{n}\left( \theta \right)
\right) =\frac{k!}{\left( n\theta \right) _{k}}\prod_{m=1}^{n}\frac{\left(
\theta \right) _{k_{m}}}{k_{m}!},  \label{e1}
\end{equation}
which is the Dirichlet-multinomial distribution, with $\mathbf{E}\left(
K_{n,k}\left( m\right) \right) =k/n$. We shall put $\mathbf{K}_{n,k}\overset{%
d}{\sim }D_{n,k}\left( \theta \right) $.

The sampling from $\mathbf{S}_{n}\left( \theta \right) \overset{d}{\sim }$ $%
D_{n}\left( \theta \right) $ formula (\ref{e1}) coincides with the one (\ref
{f6d}) obtained while sampling from a discrete species abundance model $%
\mathbf{\xi }_{n}$ with negative binomial distributions. The $*-$limit of
this Dirichlet model is known to lead to the Ewens sampling formulae which
are particular incarnation of (\ref{f15a}) and (\ref{f19a}) with $\phi
_{i}=\left( i-1\right) !$ and $B_{k,p}\left( \phi _{\bullet }\right)
=s_{k,p}.$ See \cite{King} and \cite{Kingman}.\newline

It is worthwhile exploring if this remarkable property (or maybe a weaker
one) propagates to sampling from other discrete species abundance model.

\section{Sampling problems from a special CP class}

We shall now exhibit a sub-class of CP models whose statistical properties
are very similar to the ones developed in the latter Section for the
Dirichlet model.

\subsection{Sampling from a special CP class}

Let us first define the class of $\phi $ we will be interested in.

\textbf{The special class} $\mathcal{S}$.

We first recall that a function $h\left( x\right) $ defined on some interval 
$x\in \left( -\infty ,x_{0}\right) $ is absolutely monotone on some open
interval $I\subseteq \left( -\infty ,x_{0}\right) $ if it is $C^{\infty }$
with $h^{\left( n\right) }\left( x\right) \geq 0$ for all $n\geq 0$ and $%
x\in I.$

We shall consider the following special class model

\begin{definition}
Suppose that $\phi \left( x\right) $ (with $\phi _{1}>0$ and $\phi _{m}\geq
0 $, $m\geq 2$) as from (\ref{f0}), is defined (finite) on the unbounded
half-domain $x\in \left( -\infty ,x_{0}\right) $ with $0<x_{0}\leq \infty $
and that $\phi ^{\prime }\left( x\right) $ is absolutely monotone for all $%
x\in \left( -\infty ,x_{0}\right) $. If this is the case, we shall put $\phi
\in \mathcal{S}$. If $\phi \in \mathcal{S}$, $Z_{\theta }\left( x\right)
=\exp \left( \theta \phi \left( x\right) \right) $ is also defined on $x\in
\left( -\infty ,x_{0}\right) $ and absolutely monotone there.
\end{definition}

$\bullet $ Examples of $\phi \in \mathcal{S}$ are $x$, $e^{x}-1$ (Bell)$,$ $%
-\log \left( 1-x\right) $, $\left( 1-x\right) ^{-\alpha }-1,$ $\alpha >0$
and $1-\left( 1-x\right) ^{\alpha }$, $\alpha \in \left( 0,1\right) $.

$\bullet $ Examples of $\phi \notin \mathcal{S}$ are polynomials with
positive coefficients $\sum_{l=1}^{d}c_{l}x^{l}$ ($d\geq 2$), $xe^{x},$ $%
\sinh \left( x\right) $, $\cosh \left( x\right) -1$ and $\tan \left(
x\right) $. Although the latter $\phi $'s can be expanded as in (\ref{f0})
and all have non-negative Taylor coefficients $\phi _{m}$ ($\phi _{1}>0$),
the corresponding $\phi ^{\prime }\left( x\right) $ are not absolutely
monotone on $\left( -\infty ,x_{0}\right) $ although they are of course on $%
\left( 0,x_{0}\right) $.\newline

\textbf{Remarks and properties:}

- If $\phi \in \mathcal{S}$, so does clearly $\widetilde{\phi }\left(
x\right) :=a\phi \left( bx\right) $ for all $a,b>0.$ We can check that: $%
B_{k,p}\left( \widetilde{\phi }_{\bullet }\right) =a^{p}b^{k}B_{k,p}\left(
\phi _{\bullet }\right) .$

- If $\phi ^{1},\phi ^{2}\in \mathcal{S}$, then $\phi ^{1}+\phi ^{2}\in 
\mathcal{S}$ and the composition $\phi ^{1}\circ \phi ^{2}\in \mathcal{S}$.
This allows to produce a lot of new examples of $\phi $'s in $\mathcal{S}$
from the ones already introduced. For instance because $\phi ^{1}=\left(
1-x\right) ^{-\alpha }-1$ and $\phi ^{2}=1-\left( 1-x\right) ^{\alpha }$
both belong to $\mathcal{S}$, would $\alpha \in \left( 0,1\right) ,$ $\phi
^{1}+\phi ^{2}=2\sinh \left( -\alpha \log \left( 1-x\right) \right) $
belongs to $\mathcal{S}\emph{,}$ together with $\phi ^{1}\circ \phi
^{2}=\left( 1-x\right) ^{-\alpha ^{2}}-1$ and $\phi ^{2}\circ \phi
^{1}=1-\left( 2-\left( 1-x\right) ^{-\alpha }\right) ^{\alpha }.$

- If $\phi ^{1},\phi ^{2}\in \mathcal{S}$, the product $\phi :=\phi
^{1}\cdot \phi ^{2}\notin $ (in the first place because $\phi _{1}=0$)$.$
The Taylor coefficients $\phi _{m}$ of $\phi $ are 
\begin{equation*}
\phi _{m}=\sum_{l=1}^{m-1}\binom{m}{l}\phi _{l}^{1}\phi _{m-l}^{2}=\left(
\phi ^{1}*\phi ^{2}\right) _{m}\text{, }m\geq 2
\end{equation*}
and the $\phi _{m}$ do not necessarily form a log-convex sequence, even
though $\phi _{m}^{1},$ $\phi _{m}^{2}$, $m\geq 1,$ would be log-convex
themselves. This is not in contradiction with the Davenport and P\`{o}lya
theorem \cite{DP} stating that the binomial convolution of two log-convex
sequences is log-convex because the $\phi ^{1},\phi ^{2}$ sequences here
have no constant terms: $\phi _{0}^{1}=\phi _{0}^{2}=0$ (resulting in $\phi
_{1}=0$)$.$ The reason why, when $\phi \left( x\right) \in \mathcal{S}$,
log-convexity of the sequences $\left( \phi _{m}\right) _{m\geq 1}$ pops in
is (see \cite{Berg} and \cite{Schill}):\newline

\begin{proposition}
When $\phi \in \mathcal{S}$, the function $h\left( x\right) :=\phi ^{\prime
}\left( -x\right) $ is completely monotone on the domain $x\in \left(
-x_{0},\infty \right) ,$meaning it is $C^{\infty }$ with $\left( -1\right)
^{n}h^{\left( n\right) }\left( x\right) \geq 0$ for all $n\geq 0$ and $x\in
\left( -x_{0},\infty \right) .$ So (from Bernstein theorem \cite{Bern}), $%
h\left( x\right) $ is the Laplace-Stieltjes transform (LST) of some finite
non-negative measure $\mu $ on $\left[ 0,+\infty \right) :$ $h\left(
x\right) =\int_{0}^{\infty }e^{-xt}\mu \left( dt\right) .$ We have 
\begin{equation*}
h\left( x\right) =\sum_{m\geq 0}\frac{\phi _{m+1}}{m!}\left( -x\right) ^{m}
\end{equation*}
and so $\phi _{m+1}$ is the $m^{\text{th}}$ moment of $\mu ,$ with finite
total mass $\phi _{1}.$ By the Cauchy-Schwarz inequality, for all $m\geq 2,$ 
$\phi _{m+1}\phi _{m-1}\geq \phi _{m}^{2}$, showing that when $\phi \in 
\mathcal{S},$ $\left( \phi _{m}\right) _{m\geq 1}$ is a log-convex sequence.
Upon shifting, $\left( \phi _{m}\right) _{m\geq 1}$ is the moment sequence
of some non-negative measure $\pi \left( dt\right) :=t^{-1}\mu \left(
dt\right) .$
\end{proposition}

Let us now consider $Z_{\theta }\left( -x\right) =e^{\theta \phi \left(
-x\right) }=:e^{-\theta \psi \left( x\right) }$, with 
\begin{equation*}
\psi \left( x\right) :=-\phi \left( -x\right) \text{, }x>-x_{0}.
\end{equation*}

\begin{proposition}
When $\phi \in \mathcal{S}$, it holds that $\psi ^{\prime }\left( x\right)
=h\left( x\right) =\int_{0}^{\infty }e^{-xt}t\pi \left( dt\right) $ is
completely monotone, so $Z_{\theta }\left( -x\right) =e^{-\theta \psi \left(
x\right) }$ is the LST of some infinitely divisible random variable (or
subordinator process) $Y_{\theta }$ on $\left[ 0,+\infty \right) ,$ whose
integral moments are all finite. The coefficients $\left( \phi _{m}\right)
_{m\geq 1}$ are the cumulants of $Y_{\theta }$. The function $\psi $ is the
Laplace exponent of $Y_{\theta }$ with $\psi \left( x\right)
=cx+\int_{0}^{\infty }\left( 1-e^{-xt}\right) \pi \left( dt\right) $ for
some $c\geq 0$ and some positive L\'{e}vy measure $\pi \left( dt\right) $ on 
$\left( 0,\infty \right) ,$ integrating $1\wedge t$ \cite{Steu}. Therefore,
when $\phi \in \mathcal{S}$, 
\begin{equation*}
Z_{\theta }\left( -x\right) =\mathbf{E}\left( e^{-xY_{\theta }}\right)
=e^{-\theta \psi \left( x\right) }=1+\sum_{k\geq 1}\frac{\left( -x\right)
^{k}}{k!}\sigma _{k}\left( \theta \right) ,
\end{equation*}
with $\left( \sigma _{k}\left( \theta \right) ,\text{ }k\geq 0\right) $
being the Stieltjes moment sequence of $Y_{\theta }:$ $\sigma _{k}\left(
\theta \right) =\mathbf{E}\left( Y_{\theta }^{k}\right) .$ Thus, when $\phi
\in \mathcal{S}$, for all $\theta >0,$ $\left( \sigma _{k}\left( \theta
\right) \right) _{k\geq 0}$ forms a $k-$log-convex sequence and for all $%
k\geq 1,$ all $\theta >0:$ $\sigma _{k+1}\left( \theta \right) \sigma
_{k-1}\left( \theta \right) \geq \sigma _{k}\left( \theta \right) ^{2}.$

Since $\mathbf{E}\left( e^{-x\overline{Y}_{n,\theta }}\right) =e^{-n\theta
\psi \left( x\right) },$ $\sigma _{k}\left( n\theta \right) $ is also the $%
k^{\text{th}}$ moment of the sum $\overline{Y}_{n,\theta }:=Y_{1,\theta
}+...+Y_{n,\theta }$ of $n$ iid terms $Y_{m,\theta }:=Y_{m\theta }-Y_{\left(
m-1\right) \theta }$. So, $\sigma _{k}\left( n\theta \right) =\mathbf{E}%
\left( \overline{Y}_{n,\theta }^{k}\right) =\mathbf{E}\left( Y_{n\theta
}^{k}\right) .$\newline
\end{proposition}

Note finally that taking $Z_{\theta }\left( x\right) =Z_{\theta }^{1}\left(
x\right) Z_{\theta }^{2}\left( x\right) $ where $Z_{\theta }^{i}\left(
x\right) =e^{\theta \phi _{i}\left( x\right) }$ for two $\phi _{i}$ in $%
\mathcal{S}$, with $\sigma _{k}^{i}\left( \theta \right) $ defined by $%
Z_{\theta }^{i}\left( x\right) =1+\sum_{k\geq 1}\frac{x^{k}}{k!}\sigma
_{k}^{i}\left( \theta \right) $, two $k-$log-convex sequences, the sequence $%
\sigma _{k}\left( \theta \right) $ defined by $Z_{\theta }\left( x\right)
=1+\sum_{k\geq 1}\frac{x^{k}}{k!}\sigma _{k}\left( \theta \right) $ obeys 
\begin{equation*}
\sigma _{k}\left( \theta \right) =\sum_{l=0}^{k}\binom{k}{l}\sigma
_{l}^{1}\left( \theta \right) \sigma _{k-l}^{2}\left( \theta \right) =\left(
\sigma ^{1}\left( \theta \right) *\sigma ^{2}\left( \theta \right) \right)
_{k}\text{, }k\geq 0,
\end{equation*}
and is $k-$log-convex by Davenport and P\`{o}lya theorem, as a binomial
convolution of two log-convex sequences.

\textbf{Sampling from } $\mathbf{\xi }_{n}$\textbf{\ when }$\phi \in 
\mathcal{S}$.

Assume $\phi \in \mathcal{S}$ and consider the sampling problem from $%
\mathbf{\xi }_{n}$, where $\xi $ is constructed as in Section $2$ from $\phi 
$, but now for $\phi \in \mathcal{S}$. Note that in this case 
\begin{equation*}
\mathbf{E}\left( u^{\xi }\right) =e^{\theta \left[ \phi \left( xu\right)
-\phi \left( x\right) \right] }=e^{-\theta \left[ \psi \left( -xu\right)
-\psi \left( -x\right) \right] }.
\end{equation*}
In a general sampling problem from $\mathbf{\xi }_{n}$, the joint
probability generating function of $\mathbf{K}_{n,k}$ was given by (\ref{f7}%
). From (\ref{f6}) and making use of $\phi \in \mathcal{S}$, from
Proposition $10$, we have 
\begin{equation}
\mathbf{P}\left( \mathbf{K}_{n,k}=\mathbf{k}_{n}\right) =\frac{k!}{\sigma
_{k}\left( n\theta \right) }\prod_{m=1}^{n}\frac{\sigma _{k_{m}}\left(
\theta \right) }{k_{m}!}=\binom{k}{k_{1}...k_{n}}\frac{\prod_{m=1}^{n}%
\mathbf{E}\left( Y_{m,\theta }^{k_{m}}\right) }{\mathbf{E}\left( \overline{Y}%
_{n,\theta }^{k}\right) },  \label{g3a}
\end{equation}
\textbf{Remark:} Because (\ref{g3a}) does not depend on the common mean of
the $Y_{m,\theta }$'s, we can as well define the reduced (iid) random
variables with mean $1:$ $X_{m,\theta }:=Y_{m,\theta }/\left( \theta \phi
_{1}\right) ,$ $m=1,...,n$ and $\overline{X}_{n,\theta
}:=\sum_{m=1}^{n}X_{m,\theta }.$ Then, with $S_{m,\theta }:=X_{m,\theta }/%
\overline{X}_{n,\theta }$, $m=1,...,n$ defining a random partition $\mathbf{S%
}_{n}\left( \theta \right) =\left( S_{1,\theta },...,S_{n,\theta }\right) $
of unity into $n$ exchangeable (mean $1/n$) parts 
\begin{equation}
\mathbf{P}\left( \mathbf{K}_{n,k}=\mathbf{k}_{n}\right) =\binom{k}{%
k_{1}...k_{n}}\frac{\prod_{m=1}^{n}\mathbf{E}\left( X_{m,\theta
}^{k_{m}}\right) }{\mathbf{E}\left( \overline{X}_{n,\theta }^{k}\right) }
\label{g4}
\end{equation}
\begin{equation*}
=\binom{k}{k_{1}...k_{n}}\frac{\prod_{m=1}^{n}\mathbf{E}\left( \overline{X}%
_{n,\theta }^{k_{m}}S_{m,\theta }^{k_{m}}\right) }{\mathbf{E}\left( 
\overline{X}_{n,\theta }^{k}\right) }=\binom{k}{k_{1}...k_{n}}\frac{\mathbf{E%
}\left( \overline{X}_{n,\theta }^{k}\prod_{m=1}^{n}S_{m,\theta
}^{k_{m}}\right) }{\mathbf{E}\left( \overline{X}_{n,\theta }^{k}\right) }
\end{equation*}
as well. The latter expression is identified to an occupancy distribution
arising from sampling from the random partition of unity $\mathbf{S}%
_{n}\left( \theta \right) $ but now biased by the total length $\overline{X}%
_{n,\theta }.$ In the occupancy distribution (\ref{g4}) indeed, realizations
of $\left( X_{m,\theta }\right) _{m=1}^{n}$ giving rise to large values of
the sum $\overline{X}_{n,\theta }$ are favored, compared to the ``unbiased''
multinomial one, say $\mathbf{Q}\left( \mathbf{K}_{n,k}=\mathbf{k}%
_{n}\right) :=\binom{k}{k_{1}...k_{n}}\prod_{m=1}^{n}\mathbf{E}\left(
S_{m,\theta }^{k_{m}}\right) ,$ based on the same $\mathbf{S}_{n}\left(
\theta \right) $.

Would $\overline{X}_{n,\theta }$ be independent of $S_{m,\theta
}=X_{m,\theta }/\overline{X}_{n,\theta }$, $m=1,...,n,$ (the only possible
way, by Lukacs' criterion, to have this is when $\mathbf{S}_{n}\left( \theta
\right) $ has Dirichlet$\left( \theta \right) $ distribution, \cite{Holst}),
the latter expression boils down to the usual sampling one 
\begin{equation*}
\mathbf{P}\left( \mathbf{K}_{n,k}=\mathbf{k}_{n}\right) =\binom{k}{%
k_{1}...k_{n}}\mathbf{E}\left( \prod_{m=1}^{n}S_{m,\theta }^{k_{m}}\right) =%
\mathbf{Q}\left( \mathbf{K}_{n,k}=\mathbf{k}_{n}\right) .
\end{equation*}
Alternatively, from (\ref{g4}), the joint pgf of $\mathbf{K}_{n,k}$ also
reads 
\begin{equation*}
\mathbf{E}\left[ \prod_{m=1}^{n}u_{m}^{K_{n,k}\left( m\right) }\right] =%
\frac{\mathbf{E}\left[ \left( \sum_{m=1}^{n}u_{m}X_{m,\theta }\right)
^{k}\right] }{\mathbf{E}\left( \overline{X}_{n,\theta }^{k}\right) }=\frac{%
\mathbf{E}\left[ \overline{X}_{n,\theta }^{k}\left(
\sum_{m=1}^{n}u_{m}S_{m,\theta }\right) ^{k}\right] }{\mathbf{E}\left( 
\overline{X}_{n,\theta }^{k}\right) }.
\end{equation*}
\newline
Its computation is thus amenable to the normalized $k^{\text{th}}$ moment of
the weighted sum $\sum_{1}^{n}u_{m}X_{m,\theta }$ of iid mean $1$ infinitely
divisible random variables with LST $\mathbf{E}\left( e^{-xX_{\theta
}}\right) =e^{\theta \phi \left( -x/\left( \theta \phi _{1}\right) \right)
}=e^{-\theta \psi \left( x/\left( \theta \phi _{1}\right) \right) }$ and
moments $\mathbf{E}\left( X_{\theta }^{k}\right) =\sigma _{k}\left( \theta
\right) /\left( \theta \phi _{1}\right) ^{k},k\geq 1.$ Unlike $\left(
Y_{\theta };\theta \geq 0\right) $, the process $\left( X_{\theta };\theta
\geq 0\right) $ is not a L\'{e}vy process.

Note also that with $\mathbf{k}_{p}:=\left( k_{1},...,k_{p}\right) \in \Bbb{N%
}^{p}$ obeying $\left| \mathbf{k}_{p}\right| =k$

\begin{equation*}
\mathbf{P}\left( \widehat{K}_{n,k}\left( 1\right) =k_{1},...,\widehat{K}%
_{n,k}\left( p\right) =k_{p};P_{n,k}=p\right) =
\end{equation*}
\begin{equation*}
\binom{n}{p}\binom{k}{k_{1}...k_{p}}\frac{\prod_{q=1}^{p}\mathbf{E}\left( 
\overline{X}_{n,\theta }^{k_{q}}S_{q,\theta }^{k_{q}}\right) }{\mathbf{E}%
\left( \overline{X}_{n,\theta }^{k}\right) }=\binom{n}{p}\binom{k}{%
k_{1}...k_{p}}\frac{\mathbf{E}\left( \overline{X}_{n,\theta
}^{k}\prod_{q=1}^{p}S_{q,\theta }^{k_{q}}\right) }{\mathbf{E}\left( 
\overline{X}_{n,\theta }^{k}\right) }
\end{equation*}
is the joint probability that there are $p\in \left[ n\right] $ non-empty
boxes and that $\left( k_{1},...,k_{p}\right) $ are the respective
occupancies of the $p$ filled boxes, labeled in arbitrary order. Again 
\begin{eqnarray*}
\mathcal{P}_{k,p}^{\left( n\right) } &:&=\binom{n}{p}\sum_{\mathbf{k}_{p}\in 
\Bbb{N}^{p}:\left| \mathbf{k}_{p}\right| =k}\binom{k}{k_{1}...k_{p}}\frac{%
\prod_{q=1}^{p}\mathbf{E}\left( \overline{X}_{n,\theta }^{k_{q}}S_{q,\theta
}^{k_{q}}\right) }{\mathbf{E}\left( \overline{X}_{n,\theta }^{k}\right) } \\
&=&\binom{n}{p}\sum_{\mathbf{k}_{p}\in \Bbb{N}^{p}:\left| \mathbf{k}%
_{p}\right| =k}\binom{k}{k_{1}...k_{p}}\frac{\mathbf{E}\left( \overline{X}%
_{n,\theta }^{k}\prod_{q=1}^{p}S_{q,\theta }^{k_{q}}\right) }{\mathbf{E}%
\left( \overline{X}_{n,\theta }^{k}\right) }
\end{eqnarray*}
is the probability that in a $k-$sample from $n$ species with abundance $%
\mathbf{\xi }_{n}$ in the special class $\mathcal{S}$, the exact number of
distinct visited species is $p.$

To summarize, we conclude

\begin{proposition}
When $\phi \in \mathcal{S}$ and when the discrete species abundance model $%
\mathbf{\xi }_{n}$ is built on $\phi $, its occupancy distribution (\ref{f6}%
) can alternatively be given the interpretation of an occupancy distribution
(\ref{g4}) arising from sampling from the random partition of unity $\mathbf{%
S}_{n}\left( \theta \right) $ but biased by the total length $\overline{X}%
_{n,\theta }$ appearing in the normalization of $S_{m,\theta }:=X_{m,\theta
}/\overline{X}_{n,\theta }.$ The positive random variable $X_{\theta }%
\overset{d}{=}X_{1,\theta }$ is infinitely divisible. The correspondence
between $\xi $ and (mean $1$) $X_{\theta }$ is: 
\begin{equation*}
\mathbf{E}\left[ u^{\xi }\right] =e^{-\theta \phi \left( x\right) \left( 1-%
\frac{\phi \left( xu\right) }{\phi \left( x\right) }\right) }\text{ and }%
\mathbf{E}\left( e^{-xX_{\theta }}\right) =e^{\theta \phi \left( -x/\left(
\theta \phi _{1}\right) \right) }=e^{-\theta \psi \left( x/\left( \theta
\phi _{1}\right) \right) }.
\end{equation*}
\end{proposition}

Note finally that $\psi \left( x/\left( \theta \phi _{1}\right) \right) $
being the Laplace exponent of $X_{\theta }:$%
\begin{equation*}
\mathbf{E}\left( e^{-xX_{\theta }}\right) =e^{-\theta \int_{0}^{\infty
}\left( 1-e^{-xt}\right) \pi _{\theta }\left( dt\right) },
\end{equation*}
where the L\'{e}vy measure $\pi _{\theta }\left( dt\right) $ integrates $%
1\wedge t.$ The measure $t\pi _{\theta }\left( dt\right) $ is a finite
positive measure with all finite $m-$moments: $\int_{0}^{\infty }t^{m}t\pi
_{\theta }\left( dt\right) =\phi _{m+1}/\left( \theta \phi _{1}\right)
^{m+1} $, $m\geq 0.$ So $\left( \left( \theta \phi _{1}\right) ^{-m}\phi
_{m}\right) _{m\geq 1}$ is the moment sequence of $\pi _{\theta }\left(
dt\right) $.

With $S_{1,\theta }:=X_{1,\theta }/\overline{X}_{n,\theta }$, define finally 
$\mu _{k}:=\mathbf{E}\left[ S_{1,\theta }^{k}\right] $, $k\geq 1,$\ the
sequence of the moments of $S_{1,\theta }$.\newline

\textbf{Examples.} Examples of admissible $\phi \in \mathcal{S}$ were $-\log
\left( 1-x\right) $, $\left( 1-x\right) ^{-\alpha }-1,$ $\alpha >0$ and $%
1-\left( 1-x\right) ^{\alpha }$, $\alpha \in \left( 0,1\right) $.

The LST $\mathbf{E}\left( e^{-xX_{\theta }}\right) $ of $X_{\theta }$ in
each case is $\left( 1+x/\theta \right) ^{-\theta }$, $\exp \left[ -\theta
\left( 1-\left( 1+\frac{x}{\alpha \theta }\right) ^{-\alpha }\right) \right] 
$ and $\exp \left[ -\theta \left( \left( 1+\frac{x}{\alpha \theta }\right)
^{\alpha }-1\right) \right] $ corresponding respectively to a Gamma$\left(
\theta ,\theta \right) $ distribution, a compound Poisson sum of iid gamma$%
\left( \alpha ,\alpha \theta \right) $ random variables and an exponentially
damped stable$\left( \theta ,\alpha \right) $. For this last case, let $%
\Sigma >0$ be a stable$\left( \theta ,\alpha \right) $ random variable i.e.
with LST $\mathbf{E}\left( e^{-x\Sigma }\right) :=\exp \left[ -\theta
x^{\alpha }\right] $, $x\geq 0.$ Let $f_{\Sigma }$ be its density. Define a
random variable $Y_{\theta }$ with damped density $f_{Y_{\theta }}\left(
t\right) =\frac{1}{\mathbf{E}\left( e^{-\Sigma }\right) }e^{-t}f_{\Sigma
}\left( t\right) $, $t>0.$ Its LST is $\mathbf{E}\left( e^{-xY_{\theta
}}\right) =\mathbf{E}\left( e^{-\left( x+1\right) \Sigma }\right) /\mathbf{E}%
\left( e^{-\Sigma }\right) =\exp -\theta \left[ \left( 1+x\right) ^{\alpha
}-1\right] $. Upon scaling $Y_{\theta }$, $X_{\theta }:=Y_{\theta }/\left(
\theta \alpha \right) $ is mean $1$. In the sampling context, the last
example was recently considered in (\cite{Eng}, \cite{Engen}, \cite{Hosh1}
and \cite{Hosh2}). They were named the generalized inverse Gaussian or Engen
models. $\diamondsuit $\newline

\textbf{Remark.} in the degenerate case, $\phi \left( x\right) =x$, $%
X_{\theta }$ is purely atomic with $X_{\theta }\overset{d}{\sim }\delta _{1}$%
. The LST of $X_{\theta }$ can be obtained from the one of the first gamma$%
\left( \theta ,\theta \right) $ example: $\mathbf{E}\left( e^{-xX_{\theta
}}\right) =\left( 1+x/\theta \right) ^{-\theta }$ as $\theta \rightarrow
\infty $. In this very particular (admissible) case, $\mathbf{S}_{n}=\left(
1/n,...,1/n\right) $ is the uniform deterministic partition of unity (the
Maxwell-Boltzmann case). $\diamondsuit $\newline

\subsection{The *-limit}

We now come back to the $*-$limit.

Let $\phi \in \mathcal{S}$. With $\gamma >0$, let $\left( Y_{\gamma }\right)
_{\gamma \geq 0}$ be a subordinator with $Y_{0}=0$ and LST 
\begin{equation*}
\mathbf{E}\left( e^{-xY_{\gamma }}\right) =e^{-\gamma \psi \left( x\right) },%
\text{ }\psi \left( x\right) =-\phi \left( -x\right) .
\end{equation*}
Under our assumptions on $\phi $, $\mathbf{E}\left( Y_{\gamma }\right)
=\gamma \phi _{1}<\infty .$ Then the Laplace exponent $\psi $ reads 
\begin{equation}
\psi \left( x\right) =\int_{0}^{\infty }\left( 1-e^{-xt}\right) \pi \left(
dt\right) ,  \label{h1}
\end{equation}
for some positive L\'{e}vy measure $\pi $ on $\left( 0,\infty \right) ,$
integrating $1\wedge t,$ \cite{Bertoin}$.$ Let $\overline{\pi }\left(
t\right) :=\int_{t}^{\infty }\pi \left( ds\right) $ be the tail function of $%
\pi $ and assume $\overline{\pi }\left( t\right) \rightarrow \infty $ as $%
t\rightarrow 0$(\footnote{%
If $\overline{\pi }$ has a finite limit, the random partition of unity
defined in (\ref{h5}) is finite with a random Poisson number of pieces (see
Example $\left( iii\right) $ below). The corresponding subordinator has an
atom at point $\gamma =0$ with positive probability. This case deserves a
special treatment.}). Then 
\begin{equation}
Y_{\gamma }=\sum_{k\geq 1}\overline{\pi }^{-1}\left( \Gamma _{k}/\gamma
\right)  \label{h2}
\end{equation}
where $\left( \Gamma _{k}\right) _{k\geq 1}$ are the points of a standard
Poisson Point Process (PPP) on $\left( 0,\infty \right) $ with intensity $1.$
The random variables 
\begin{equation*}
\Delta _{\left( k\right) }\left( \gamma \right) :=\overline{\pi }^{-1}\left(
\Gamma _{k}/\gamma \right)
\end{equation*}
with $\Delta _{\left( 1\right) }\left( \gamma \right) \geq \Delta _{\left(
2\right) }\left( \gamma \right) \geq ...$ constitute the ranked jumps'
heights of the subordinator $Y_{\gamma }$ (they are countably many, with $0$
as a limit point). They form a PPP on the half-line with intensity $\gamma
\pi \left( dt\right) ,$ and the law of $\Delta _{\left( k\right) }\left(
\gamma \right) $ can easily be computed to be \cite{Bertoin} 
\begin{equation}
\mathbf{P}\left( \Delta _{\left( k\right) }\left( \gamma \right) \in
dt\right) =\frac{\gamma ^{k}\overline{\pi }\left( t\right) ^{k-1}}{\left(
k-1\right) !}e^{-\gamma \overline{\pi }\left( t\right) }\pi \left( dt\right)
.  \label{h3}
\end{equation}
By Campbell formula (see \cite{Neveu}, \cite{Kingman}), for all measurable
function $g$ for which $\int_{0}^{\infty }\left( 1-e^{-xg\left( t\right)
}\right) \pi \left( dt\right) <\infty ,$ we have 
\begin{equation*}
\mathbf{E}\left( \exp \left\{ -x\sum_{k\geq 1}g\left( \overline{\pi }%
^{-1}\left( \Gamma _{k}/\gamma \right) \right) \right\} \right) =\exp
\left\{ -\gamma \int_{0}^{\infty }\left( 1-e^{-xg\left( t\right) }\right)
\pi \left( dt\right) \right\} .
\end{equation*}
Putting $g\left( t\right) =t$, $\mathbf{E}\left( e^{-xY_{\gamma }}\right)
=e^{-\gamma \psi \left( x\right) }$, showing that (\ref{h2}) holds in law.%
\newline

From the above construction, when $\pi $ has infinite mass, we can define a
random distribution on the infinite-dimensional $1-$simplex by normalizing
the ranked jumps' heights of $Y_{\gamma }$ by itself. Consider again $%
Y_{\gamma }$ and, with $\theta :=\gamma /n$, define $Y_{m,\theta
}:=Y_{m\theta }-Y_{\left( m-1\right) \theta }$, $m=1,...,n$ which are
mutually independent$.$ Then, $\overline{Y}_{n,\theta
}:=\sum_{m=1}^{n}Y_{m,\theta }=Y_{n\theta }=Y_{\gamma }.$ If we rank the $%
Y_{m,\theta }$'s, with $Y_{\left( 1\right) ,\theta }\geq ...\geq Y_{\left(
n\right) ,\theta }$(\footnote{%
If $Y_{\theta }$ has a density ($\pi $ has no atom), these inequalities are
strict.}), then, \cite{Kingm}, as $n\rightarrow \infty $, $\theta
\rightarrow 0,$ $n\theta =\gamma $%
\begin{equation}
\left( Y_{\left( 1\right) ,\theta },...,Y_{\left( n\right) ,\theta
},0,0,...\right) \overset{d}{\underset{*}{\rightarrow }}\left( \Delta
_{\left( 1\right) }\left( \gamma \right) ,\Delta _{\left( 2\right) }\left(
\gamma \right) ,...\right) .  \label{h4}
\end{equation}
Normalizing, 
\begin{equation*}
\left( Y_{\left( 1\right) ,\theta }/Y_{\gamma },...,Y_{\left( n\right)
,\theta }/Y_{\gamma },0,0,...\right) \overset{d}{\underset{*}{\rightarrow }}
\end{equation*}
\begin{equation}
\left( \Delta _{\left( 1\right) }\left( \gamma \right) /Y_{\gamma },\Delta
_{\left( 2\right) }\left( \gamma \right) /Y_{\gamma },...\right) =:\mathbf{S}%
_{\infty }\left( \gamma \right) :=\left( S_{\left( 1\right) ,\gamma
},S_{\left( 2\right) ,\gamma },...\right) ,  \label{h5}
\end{equation}
with $\mathbf{S}_{\infty }\left( \gamma \right) $ defining a random
partition of unity with infinitely many (ordered) pieces.

If $t>0$ is some (small) cutoff or threshold value, let $N_{+}\left(
t\right) :=\sum_{k\geq 1}\mathbf{I}\left( \Delta _{\left( k\right) }\left(
\gamma \right) >t\right) $ count the numbers of atoms of the partition of $%
Y_{\gamma }$ exceeding $t$. By Campbell formula 
\begin{equation*}
\mathbf{E}\left( \exp \left\{ -xN_{+}\left( t\right) \right\} \right) =\exp
\left\{ -\gamma \int_{0}^{\infty }\left( 1-e^{-x\mathbf{I}\left( s>t\right)
}\right) \pi \left( ds\right) \right\}
\end{equation*}
\begin{equation}
=\exp \left\{ -\gamma \overline{\pi }\left( t\right) \left( 1-e^{-x}\right)
\right\}  \label{eq12}
\end{equation}
is the full LST of $N_{+}\left( t\right) $. This shows that $N_{+}\left(
t\right) $ is Poisson distributed with mean $\gamma \overline{\pi }\left(
t\right) $. Recalling $\overline{\pi }\left( t\right) \underset{t\rightarrow
0}{\rightarrow }\infty $, the law of large numbers gives 
\begin{equation}
N_{+}\left( t\right) /\overline{\pi }\left( t\right) \overset{a.s.}{%
\rightarrow }\gamma \text{, as }t\rightarrow 0.  \label{eq13}
\end{equation}
The fact that $N_{+}\left( t\right) $ is Poisson may be also checked as
follows. We have $N_{+}\left( t\right) =\inf \left( k\geq 1:\Delta _{\left(
k\right) }\left( \gamma \right) \leq t\right) -1$ and $\mathbf{P}\left(
N_{+}\left( t\right) \geq k\right) =\mathbf{P}\left( \Delta _{\left(
k\right) }\left( \gamma \right) >t\right) =\mathbf{P}\left( \Gamma _{k}\leq
\gamma \overline{\pi }\left( t\right) \right) =e^{-\gamma \overline{\pi }%
\left( t\right) }\sum_{l\geq k}\frac{\left[ \gamma \overline{\pi }\left(
t\right) \right] ^{l}}{l!}.$ So $N_{+}\left( t\right) $ is Poisson with mean 
$\gamma \overline{\pi }\left( t\right) $.

Because also, by the strong law of large numbers, $\Gamma _{k}/k\rightarrow
1 $ a.s. as $k\rightarrow \infty $, recalling $\Gamma _{k}=\gamma \overline{%
\pi }\left( Y_{\gamma }S_{\left( k\right) ,\gamma }\right) $, we get 
\begin{equation*}
\gamma \overline{\pi }\left( Y_{\gamma }S_{\left( k\right) ,\gamma }\right)
/k\rightarrow 1\text{ }a.s.\text{ as }k\rightarrow \infty .
\end{equation*}
From the behavior of $\overline{\pi }\left( t\right) $ near $t=0$, the decay
rate of $S_{\left( k\right) ,\gamma }$ to $0$ as $k\rightarrow \infty $
follows.

Sampling from $S_{m,\theta }:=Y_{m,\theta }/Y_{\gamma }$, $m=1,...,n.$
Define as in (\ref{g3a}) a biased sampling procedure for which ($\left| 
\mathbf{k}_{n}\right| =k$) 
\begin{equation}
\mathbf{P}\left( \mathbf{K}_{n,k}=\mathbf{k}_{n}\right) =\binom{k}{%
k_{1}...k_{n}}\frac{\mathbf{E}\left( Y_{\gamma
}^{k}\prod_{m=1}^{n}S_{m,\theta }^{k_{m}}\right) }{\mathbf{E}\left(
Y_{\gamma }^{k}\right) }.  \label{h6}
\end{equation}
Recall that this biased procedure is not the standard sampling one from a $k$
uniform throw on $S_{m,\theta }$, $m=1,...,n,$ obtained while counting the
number of uniform hits within each $S_{m,\theta }$. Indeed, would the latter
sampling model hold, instead of (\ref{h6}), one would rather expect the
strict multinomial occupancy distribution 
\begin{equation*}
\mathbf{Q}\left( \mathbf{K}_{n,k}=\mathbf{k}_{n}\right) =\binom{k}{%
k_{1}...k_{n}}\mathbf{E}\left( \prod_{m=1}^{n}\left( Y_{m,\theta }/Y_{\gamma
}\right) ^{k_{m}}\right) ,
\end{equation*}
and in general, we have $\mathbf{Q}\left( \mathbf{K}_{n,k}=\mathbf{k}%
_{n}\right) \neq \mathbf{P}\left( \mathbf{K}_{n,k}=\mathbf{k}_{n}\right) .$
According to (\ref{h6}), the joint pgf of $\mathbf{K}_{n,k}$ is 
\begin{equation*}
\mathbf{E}\left( \prod_{m=1}^{n}u_{m}^{K_{n,k}\left( m\right) }\right) =%
\frac{1}{\mathbf{E}\left( Y_{\gamma }^{k}\right) }\sum_{\mathbf{k}_{n}\in 
\Bbb{N}_{0}^{n}:\text{ }\left| \mathbf{k}_{n}\right| =k}\binom{k}{%
k_{1}...k_{n}}\prod_{m=1}^{n}u_{m}^{k_{m}}\mathbf{E}\left(
\prod_{m=1}^{n}Y_{m,\theta }^{k_{m}}\right)
\end{equation*}
\begin{equation}
=\frac{\mathbf{E}\left[ \left( \sum_{m=1}^{n}u_{m}Y_{m,\theta }\right)
^{k}\right] }{\mathbf{E}\left( Y_{\gamma }^{k}\right) },  \label{h7}
\end{equation}
which is akin to (\ref{g3a}).

Biased sampling from $\mathbf{S}_{\infty }\left( \gamma \right) =\left(
S_{\left( 1\right) ,\gamma },S_{\left( 2\right) ,\gamma },...\right) $ can
also be defined whenever the sampling process amounts to draw $k$ points at
random in the unit interval partitioned according to $\mathbf{S}_{\infty
}\left( \gamma \right) $, counting the number of points in each subintervals
and when biasing some functional $f\left( S_{\left( 1\right) ,\gamma
},S_{\left( 2\right) ,\gamma },...\right) $ under concern to produce $%
\mathbf{E}^{*}\left( Y_{\gamma }^{k}f\left( S_{\left( 1\right) ,\gamma
},S_{\left( 2\right) ,\gamma },...\right) \right) /\mathbf{E}^{*}\left(
Y_{\gamma }^{k}\right) $ when averaging over $\mathbf{S}_{\infty }\left(
\gamma \right) .$

From these considerations, we can state the following results:

\begin{proposition}
Let $\gamma =n\theta $. When $\phi \in \mathcal{S},$ with $\left( \sigma
_{k}\left( \theta \right) ,\text{ }k\geq 0\right) $ the Stieltjes moment
sequence of some infinitely divisible subordinator $Y_{\gamma }$ with
Laplace exponent $\psi \left( x\right) =-\phi \left( -x\right) $, the
occupancy distributions (\ref{f6}), (\ref{f9}) and (\ref{f12}) are biased
sampling multinomial distributions from $S_{m,\theta }:=Y_{m,\theta
}/Y_{\gamma }$, $m=1,...,n$ as defined by (\ref{h6})$.$
\end{proposition}

\begin{corollary}
When $\phi \in \mathcal{S}$ and $\pi $ has infinite mass ($\phi \left(
x\right) \underset{x\rightarrow -\infty }{\rightarrow }-\infty $)$,$ the
occupancy distributions (\ref{f15}), (\ref{f15b}) and (\ref{f19}) are biased
sampling multinomial distributions from $\mathbf{S}_{\infty }\left( \gamma
\right) =\left( S_{\left( 1\right) ,\gamma },S_{\left( 2\right) ,\gamma
},...\right) $ defined in (\ref{h5}) from the subordinator $Y_{\gamma }$
with Laplace exponent $\psi \left( x\right) =-\phi \left( -x\right) .$
\end{corollary}

\textbf{Proof: }The proof follows from the previous Proposition, the fact
that (\ref{f15}) and (\ref{f19}) were obtained as weak $*-$limits of (\ref
{f9}) and (\ref{f12}), from (\ref{h5}) and from exchangeability of the $%
K_{n,k}\left( m\right) $'s. $\diamond $\newline

Let us now illustrate Corollary $13$. For instance, when $\phi \in \mathcal{S%
}$, from (\ref{f15}), 
\begin{equation*}
\mathbf{P}^{*}\left( \widehat{K}_{k}\left( 1\right) =k_{1},...,\widehat{K}%
_{k}\left( p\right) =k_{p};P_{k}=p\right) =\frac{k!}{p!}\frac{\gamma ^{p}}{%
\sigma _{k}\left( \gamma \right) }\prod_{q=1}^{p}\frac{\phi _{k_{q}}}{k_{q}!}
\end{equation*}
\begin{equation*}
=\frac{\mathbf{E}^{*}\left( Y_{\gamma }^{k}\sum_{1\leq
m_{1}<...<m_{p}}\prod_{q=1}^{p}S_{\left( m_{q}\right) ,\gamma
}^{k_{q}}\right) }{\mathbf{E}^{*}\left( Y_{\gamma }^{k}\right) }
\end{equation*}
is the probability that there are $p$ observed species, labeled in arbitrary
way, in the $k-$sample, each visited $k_{q}$ times, and that they were
obtained after biased sampling from $S_{\left( m_{1}\right) ,\gamma
}>...>S_{\left( m_{p}\right) ,\gamma }$ for any ordered sequence $1\leq
m_{1}<...<m_{p}.$

In particular, the probability that, in a biased sampling procedure from $%
\mathbf{S}_{\infty }\left( \gamma \right) $, all elements of the $k-$sample
are of the same species (whichever species it can be) is thus 
\begin{equation}
\mathbf{P}^{*}\left( \widehat{K}_{k}\left( 1\right) =k;P_{k}=1\right) =\frac{%
\mathbf{E}^{*}\left( Y_{\gamma }^{k}\sum_{m\geq 1}S_{\left( m\right) ,\gamma
}^{k}\right) }{\mathbf{E}^{*}\left( Y_{\gamma }^{k}\right) }=\gamma \frac{%
\phi _{k}}{\sigma _{k}\left( \gamma \right) }=\mathbf{E}^{*}\left(
A_{k}\left( k\right) \right) .  \label{h8}
\end{equation}
The latter identity also follows from (\ref{f15b}) with $a_{1}=...=a_{k-1}=0$%
, $a_{k}=1$ and $p=1$ (only one species visited $k$ times).

We observe that, as $\gamma \rightarrow 0$ (or $\mathbf{E}^{*}\left(
Y_{\gamma }\right) \rightarrow 0$ as well)$,$ due to $\sigma _{k}\left(
\gamma \right) \sim \gamma \phi _{k}$, this probability tends to $1$,
showing that $\gamma $ itself may be viewed as some temperature parameter
for the population with infinitely many species: the smaller $\gamma $, the
larger the probability is that any $k-$sample visits a single one species
(among which the one with largest frequency $S_{\left( 1\right) ,\gamma }$).

Similarly, the probability that all elements of the $k-$sample reveal only
two species (whichever species they can be) is 
\begin{equation*}
\sum_{l=1}^{k-1}\frac{\mathbf{E}^{*}\left( Y_{\gamma }^{k}\sum_{1\leq
m_{1}<m_{2}}S_{\left( m_{1}\right) ,\gamma }^{l}S_{\left( m_{2}\right)
,\gamma }^{k-l}\right) }{\mathbf{E}^{*}\left( Y_{\gamma }^{k}\right) }=\frac{%
1}{2}\frac{\gamma ^{2}k!}{\sigma _{k}\left( \gamma \right) }\sum_{l=1}^{k-1}%
\frac{\phi _{l}}{l!}\frac{\phi _{k-l}}{\left( k-l\right) !}=\frac{\gamma ^{2}%
}{\sigma _{k}\left( \gamma \right) }B_{k,2}\left( \phi _{\bullet }\right) .
\end{equation*}
This identity follows from (\ref{f15b}) with $a_{l}=1,$ $a_{k-l}=1,$ $%
a_{j}=0 $ if $j\neq \left\{ l,k-l\right\} $ and $p=2$ (only two species
visited, one $l$ times and the other one $k-l$ times), summing on $%
l=1,...,k-1$ and from $\phi _{k}^{*2}=2B_{k,2}\left( \phi _{\bullet }\right) 
$. More generally, if $p\leq k$, $\frac{\gamma ^{p}}{\sigma _{k}\left(
\gamma \right) }B_{k,p}\left( \phi _{\bullet }\right) $ is the probability
that all elements of the $k-$sample reveal $p$ distinct species
(consistently with (\ref{f16})), $\frac{\left( \gamma \phi _{1}\right) ^{k}}{%
\sigma _{k}\left( \gamma \right) }$ the probability that all species in the $%
k-$sample are of distinct types. When $\gamma $ is small this latter
probability is polynomially small $\sim \gamma ^{k-1}.$

Finally, the probability that only one species is visited by the $k-$sample
and that it is the $m^{\text{th}}$ more abundant one is 
\begin{equation}
\frac{\mathbf{E}^{*}\left( Y_{\gamma }^{k}S_{\left( m\right) ,\gamma
}^{k}\right) }{\mathbf{E}^{*}\left( Y_{\gamma }^{k}\right) }=\frac{\mathbf{E}%
^{*}\left( \Delta _{\left( m\right) }\left( \gamma \right) ^{k}\right) }{%
\mathbf{E}^{*}\left( Y_{\gamma }^{k}\right) }=\frac{1}{\left( m-1\right) !}%
\frac{\int_{0}^{\infty }e^{-x}x^{m-1}\overline{\pi }^{-1}\left( x/\gamma
\right) ^{k}dx}{\sigma _{k}\left( \gamma \right) }  \label{h9}
\end{equation}
\begin{equation*}
=\frac{\gamma }{\sigma _{k}\left( \gamma \right) }\frac{1}{\left( m-1\right)
!}\int_{0}^{\infty }t^{k}\left( \gamma \overline{\pi }\left( t\right)
\right) ^{m-1}e^{-\gamma \overline{\pi }\left( t\right) }\pi \left(
dt\right) ,
\end{equation*}
consistently with (\ref{h3}). Summing (\ref{h9}) over $m\geq 1$, we recover
from (\ref{h8}), that $\phi _{k}=\frac{1}{\gamma }\int_{0}^{\infty }%
\overline{\pi }^{-1}\left( x/\gamma \right) ^{k}dx=\int_{0}^{\infty
}t^{k}\pi \left( dt\right) $ is the $k^{\text{th}}$ moment of the L\'{e}vy
measure $\pi .$ In particular, the probability that only one species is
visited by the $k-$sample \emph{and} that it is the more abundant one is
(compare with (\ref{h8})) 
\begin{eqnarray*}
\frac{\mathbf{E}^{*}\left( Y_{\gamma }^{k}S_{\left( 1\right) ,\gamma
}^{k}\right) }{\mathbf{E}^{*}\left( Y_{\gamma }^{k}\right) } &=&\frac{\gamma 
}{\sigma _{k}\left( \gamma \right) }\int_{0}^{\infty }t^{k}e^{-\gamma 
\overline{\pi }\left( t\right) }\pi \left( dt\right) \\
&=&\frac{\gamma \phi _{k}}{\sigma _{k}\left( \gamma \right) }\left[ 1-\frac{1%
}{\phi _{k}}\int_{0}^{\infty }t^{k}\left( 1-e^{-\gamma \overline{\pi }\left(
t\right) }\right) \pi \left( dt\right) \right] .
\end{eqnarray*}
When $\gamma $ gets very small, this probability approaches $1$ from below,
up to an $\mathcal{O}\left( \gamma \right) $ residual term: again, $%
S_{\left( 1\right) ,\gamma }$ dominates the other smaller $S_{\left(
m\right) ,\gamma }$ and for small values of the biodiversity parameter $%
\gamma $ therefore, the species frequencies $S_{\left( m\right) ,\gamma };$ $%
m\geq 1$ turn out to be very disparate.

Similarly, from (\ref{f19}), when $\phi \in \mathcal{S}$

\begin{equation*}
\mathbf{P}^{*}\left( A_{k}\left( 1\right) =a_{1},...,A_{k}\left( k\right)
=a_{k};P_{k}=p\right) =\frac{\gamma ^{p}k!}{\sigma _{k}\left( \gamma \right) 
}\prod_{i=1}^{k}\frac{\left( \phi _{i}/i!\right) ^{a_{i}}}{a_{i}!}
\end{equation*}
\begin{equation*}
=\frac{k!}{\prod_{i\geq 1}\left( i!^{a_{i}}a_{i}!\right) }\frac{\mathbf{E}%
^{*}\left( Y_{\gamma }^{k}\sum \prod_{i\geq 1}\prod_{j=1}^{a_{i}}S_{\left(
m_{i,j}\right) ,\gamma }^{i}\right) }{\mathbf{E}^{*}\left( Y_{\gamma
}^{k}\right) }
\end{equation*}
where in the latter numerator, the unindexed sum runs over all distinct $%
\left( m_{i,j}\right) ,$ $i=1,...,k;$ $j=1,...,a_{i}$ with $\left(
a_{1},a_{2},...\right) $ satisfying $\sum_{i\geq 1}ia_{i}=k$ and $%
\sum_{i\geq 1}a_{i}=p.$

\section{Examples}

Let us supply some Examples illustrating our results.\newline

$\left( i\right) $ Take the Fisher logarithmic series model $\phi \left(
x\right) =-\log \left( 1-x\right) \in \mathcal{S}$, resulting in $\xi $
obeying a negative binomial distribution with parameters $\theta >0$ and $%
1-x\in \left( 0,1\right) $, \cite{FCW}. Here $\phi _{\bullet }=\left(
\bullet -1\right) !$. Then $Y_{\gamma }$ is a Moran subordinator with
L\'{e}vy-measure: $\pi \left( dt\right) =t^{-1}e^{-t}dt.$ The Laplace
exponent of $Y_{\gamma }$ is $\psi \left( x\right) =\log \left( 1+x\right) ,$
in accordance with $\psi \left( x\right) =-\phi \left( -x\right) .$ In that
particular case, $\left( S_{\left( 1\right) ,\gamma },S_{\left( 2\right)
,\gamma },...\right) \sim PD\left( 0,\gamma \right) $, a Poisson-Dirichlet
partition with parameter $\gamma ,$ \cite{Holst}, \cite{Feng}$.$ Because,
due to well-known properties of Gamma-distributed random variables, $%
Y_{\gamma }$ is independent of $S_{m,\theta }=Y_{m,\theta }/Y_{\gamma }$, $%
m=1,...,n$, the biased sampling distributions from $\left( S_{1,\theta
},...S_{n,\theta }\right) $ corresponds to the usual multinomial one. In
this well-known model for species frequency, $\sigma _{k}\left( \theta
\right) =\left( \theta \right) _{k}.$ So $\sigma _{k}\left( \theta \right)
\in ZR_{-}.$

Because $\overline{\pi }\left( t\right) \sim -\log t$ as $t\rightarrow 0$, $%
N_{+}\left( t\right) :=\#\left\{ k:\Delta _{\left( k\right) }\left( \gamma
\right) >t\right\} $ grows like $-\gamma \log t$ as $t\rightarrow 0$.
Besides, 
\begin{equation*}
-\log S_{\left( k\right) ,\gamma }\sim k/\gamma \text{ as }k\rightarrow
\infty
\end{equation*}
and the ordered frequencies decay exponentially fast with $k$: species with
small frequency get exponentially rare.

Assuming $\theta $ known, the Maximum Likelihood Estimator (MLE) estimator
of $n$ in the finitely many species model is given implicitly by $P=\widehat{%
n}\left( 1-\frac{\sigma _{k}\left( \left( \widehat{n}-1\right) \theta
\right) }{\sigma _{k}\left( \widehat{n}\theta \right) }\right) $, so here 
\begin{equation*}
P=\widehat{n}\left( 1-\frac{\left( \left( \widehat{n}-1\right) \theta
\right) _{k}}{\left( \widehat{n}\theta \right) _{k}}\right) .
\end{equation*}
When $\theta =1,$ this estimator is explicitly given by 
\begin{equation*}
\widehat{n}=\frac{\left( k-1\right) P}{k-P},
\end{equation*}
where, as conventional wisdom suggests, $\widehat{n}$ will be large when the
difference between $1/P$ and $1/k$ is small (new species are being
frequently discovered). The MLE estimator of $\gamma $ in the infinitely
many species model is given implicitly by $P=\widehat{\gamma }\frac{\sigma
_{k}^{\prime }\left( \widehat{\gamma }\right) }{\sigma _{k}\left( \widehat{%
\gamma }\right) }$, \cite{Tavare}, so here 
\begin{equation*}
P=\sum_{l=0}^{k-1}\frac{\widehat{\gamma }}{\widehat{\gamma }+l}.
\end{equation*}
The estimator $\widehat{\gamma }$ is biased but its bias decreases as $k$
grows. The alternative estimator $\widetilde{\gamma }=\frac{B_{k,P-1}\left(
\phi _{\bullet }\right) }{B_{k,P}\left( \phi _{\bullet }\right) }$ with $%
B_{k,p}\left( \phi _{\bullet }\right) =s_{k,p}$ is also biased and can be
computed using the recursion for third kind Stirling numbers 
\begin{equation*}
B_{k+1,p}\left( \left( \bullet -1\right) !\right) =B_{k,p-1}\left( \left(
\bullet -1\right) !\right) +kB_{k,p}\left( \left( \bullet -1\right) !\right) 
\text{.}
\end{equation*}
\newline

$\left( ii\right) $ The full two-parameters $PD\left( \alpha ,\gamma \right) 
$ defined in \cite{PitYor} can be obtained while subordinating the damped $%
\alpha -$stable subordinator (see $(iii)$ below) to an independent Moran one
with parameter $\gamma /\alpha $. And considering the normalized ranked
sizes of the subordinate jumps$:$ here, independently of this partition of
unity, $Y_{\gamma }$ again is gamma$\left( \gamma \right) $ distributed. As
shown in \cite{PitYor}, $PD\left( \alpha ,\gamma \right) $ has many
interesting properties, \cite{Pit1}, \cite{Feng}. This partition of unity
leads to a generalized (unbiased) Ewens' sampling formula called Pitman's
sampling formula, \cite{Pit3}. Connection of the two-parameters $PD\left(
\alpha ,\gamma \right) $ partition to Gibbs (EPPF) partitions and a complete
classification of EPPFs induced by the unbiased multinomial sampling from
partition of unity can be found in \cite{Ho} and \cite{GP}.\newline

$\left( iii\right) $ Take $\phi \left( x\right) =\left( 1-x\right) ^{-\alpha
}-1\in \mathcal{S}$ where $\alpha >0.$ Here $\phi _{\bullet }=\left( \alpha
\right) _{\bullet }$ resulting in $\xi $ being a Poisson sum of negative
binomial increments $\delta $. The L\'{e}vy-measure corresponding to $%
Y_{\gamma }$ is the (mean $\alpha $) Gamma$\left( \alpha ,1\right) $
probability density: $\pi \left( dt\right) =1/\Gamma \left( \alpha \right)
\cdot t^{\alpha -1}e^{-t}dt.$ The Laplace exponent of $Y_{\gamma }$ is $\psi
\left( x\right) =1-\left( 1+x\right) ^{-\alpha },$ in accordance with $\psi
\left( x\right) =-\phi \left( -x\right) .$ Because $\pi $ is integrable with
mass $1$, $Y_{\gamma }$ is a subordinator in the compound Poisson class (a
Poisson$\left( \gamma \right) $ sum of iid positive jumps with Gamma$\left(
\alpha ,1\right) $ density). For this reason, 
\begin{equation*}
Y_{\gamma }\overset{d}{=}\left[ \sum_{k=1}^{\emph{P}\left( \gamma \right) }%
\overline{\pi }^{-1}\left( U_{\left[ k\right] }\right) \right] \cdot \mathbf{%
I}\left( \emph{P}\left( \gamma \right) \geq 1\right) +0\cdot \mathbf{I}%
\left( \emph{P}\left( \gamma \right) =0\right) ,
\end{equation*}
where $\left( U_{\left[ k\right] };k\geq 1\right) $ are the ranked ($%
U_{\left[ 1\right] }<...<U_{\left[ P_{\gamma }\right] }$) points of an iid
uniform sequence $\left( U_{k};k\geq 1\right) $ on $\left( 0,1\right) $,
independent of $\emph{P}\left( \gamma \right) $ which is Poisson$\left(
\gamma \right) $ distributed$.$ Note that $Y_{\gamma }$ has an atom at $%
Y_{\gamma }=0$ with positive probability and that, would $\emph{P}\left(
\gamma \right) \geq 1$, there are finitely many (Poissonian) terms in the
L\'{e}vy decomposition of $Y_{\gamma }$. In this case, the random variables 
\begin{equation*}
\Delta _{\left( k\right) }\left( \gamma \right) :=\overline{\pi }^{-1}\left(
U_{\left[ k\right] }\right) ;\text{ }k=1,...,\emph{P}\left( \gamma \right)
\end{equation*}
with $\Delta _{\left( 1\right) }\left( \gamma \right) \geq ...\geq \Delta
_{\left( \emph{P}\left( \gamma \right) \right) }\left( \gamma \right) $
constitute the ranked (non-null) jumps' heights of the subordinator $%
Y_{\gamma }$. Considering $Y_{\gamma }$ on the event $\emph{P}\left( \gamma
\right) \geq 1$, with $\theta :=\gamma /n$, the spacings $Y_{m,\theta }$
defined by $Y_{m,\theta }:=Y_{m\theta }-Y_{\left( m-1\right) \theta }$, $%
m=1,...,n$ are non-negative and mutually independent; also $\overline{Y}%
_{n,\theta }:=\sum_{m=1}^{n}Y_{m,\theta }=Y_{n\theta }-Y_{0}=Y_{\gamma }>0.$
Normalizing the $Y_{m,\theta }$'s with $Y_{\gamma }$ defines a proper finite
random partition of unity $S_{m}$ with a random number of non-zero parts%
\emph{\ }and bias sampling (with $\pi $ finite with mass $1$) is therefore
to be understood from this partition. In the $*-$limit, its ranked
(non-null) jumps' heights are the $\Delta _{\left( k\right) }\left( \gamma
\right) $'s. Note that when $\pi $ is integrable with mass $1$, the
biodiversity parameter $\gamma $ takes on directly the interpretation of the
expected number of species in the population.

Let us come back to our case study. We first recall that for $\phi _{\bullet
}=\left( \alpha \right) _{\bullet }$ 
\begin{equation*}
B_{k+1,p}\left( \phi _{\bullet }\right) =\alpha B_{k,p-1}\left( \phi
_{\bullet }\right) +\left( k+p\alpha \right) B_{k,p}\left( \phi _{\bullet
}\right) .
\end{equation*}
When $\alpha =1$, $B_{k,p}\left( \bullet !\right) =\binom{k-1}{p-1}\frac{k!}{%
p!}$ are the Lah numbers.

Recalling also $\mathbf{P}^{*}\left( P_{k}=p\right) =\frac{\gamma ^{p}}{%
\sigma _{k}\left( \gamma \right) }B_{k,p}\left( \phi _{\bullet }\right) ,$
we get the recursion 
\begin{equation*}
\mathbf{P}^{*}\left( P_{k+1}=p\right) =\frac{\gamma ^{p}}{\sigma
_{k+1}\left( \gamma \right) }\left( \alpha B_{k,p-1}\left( \phi _{\bullet
}\right) +\left( k+p\alpha \right) B_{k,p}\left( \phi _{\bullet }\right)
\right) =
\end{equation*}
\begin{equation*}
\frac{\sigma _{k}\left( \gamma \right) }{\sigma _{k+1}\left( \gamma \right) }%
\left( \alpha \gamma \mathbf{P}^{*}\left( P_{k}=p-1\right) +\left( k+p\alpha
\right) \mathbf{P}^{*}\left( P_{k}=p\right) \right) .
\end{equation*}
This shows that the event $P_{k+1}=p$ only depends on the event $P_{k}=p-1$
(respectively $P_{k}=p$), when a new species (respectively no new species)
is being discovered as the sample size is increased by one unit$.$ And not
on further past events such as $P_{l}=p-1$ for $p-1\leq l<k.$ The transition
rates are $\lambda _{p,p+1}=\alpha \gamma \frac{\sigma _{k}\left( \gamma
\right) }{\sigma _{k+1}\left( \gamma \right) }$ (independent of $p$ but
dependent on $k$) and $\lambda _{p,p}=\left( k+p\alpha \right) \frac{\sigma
_{k}\left( \gamma \right) }{\sigma _{k+1}\left( \gamma \right) }.$ $\lambda
_{p,p+1}$ is the rate at which a new species is being discovered given $p$
of them were previously discovered in a size$-k$ sample. This suggests an
underlying sequential urn scheme, \cite{BM}, \cite{Tavare}.

The estimator $\widetilde{\gamma }=\frac{B_{k,P-1}\left( \phi _{\bullet
}\right) }{B_{k,P}\left( \phi _{\bullet }\right) }$ of $\gamma $ can easily
be evaluated numerically thanks to the three-term recurrence which $%
B_{k,p}\left( \phi _{\bullet }\right) $ fulfills. When $\alpha =1$, it is 
\begin{equation*}
\widetilde{\gamma }=\frac{P\left( P-1\right) }{k-P+1}=\frac{P}{k}\frac{1}{%
\frac{1}{P-1}-\frac{1}{k}}.
\end{equation*}
\newline

For the four following examples, an appeal to length-biased sampling
distributions from $\mathbf{S}_{\infty }\left( \gamma \right) $ is required.%
\newline

$\left( iv\right) $ With $\alpha \in \left( 0,1\right) ,$ take $\phi \left(
x\right) =1-\left( 1-x\right) ^{\alpha }\in \mathcal{S}$, resulting in $\xi $
being a Poisson sum of extended negative binomial increments $\delta $ (also
called a Poisson-Pascal random variable). Here $\phi _{1}=\alpha $, $\phi
_{m}=\alpha \left( 1-\alpha \right) _{m-1}$, $m\geq 1$ and the weight of
large clusters is smaller than in Example $\left( i\right) $ where $\phi
_{m}=\left( m-1\right) !$. We therefore expect small clusters sizes to be
enhanced. In this case, $Y_{\gamma }$ is a damped $\alpha -$stable
subordinator with L\'{e}vy-measure: $\pi \left( dt\right) =\alpha /\Gamma
\left( 1-\alpha \right) \cdot t^{-\left( \alpha +1\right) }e^{-t}dt.$ The
Laplace exponent of $Y_{\gamma }$ is $\psi \left( x\right) =\left(
1+x\right) ^{\alpha }-1,$ in accordance with $\psi \left( x\right) =-\phi
\left( -x\right) .$ The relevant subordinator is termed the generalized
gamma (see \cite{PitPK}, \cite{GP} and \cite{Ho}).

Because $\overline{\pi }\left( t\right) \sim 1/\Gamma \left( 1-\alpha
\right) \cdot t^{-\alpha }$ as $t\rightarrow 0$, $N_{+}\left( t\right)
:=\#\left\{ k:\Delta _{\left( k\right) }\left( \gamma \right) >t\right\} $
grows like $\gamma /\Gamma \left( 1-\alpha \right) \cdot t^{-\alpha }$ as $%
t\rightarrow 0$. Besides, 
\begin{equation*}
S_{\left( k\right) ,\gamma }\sim \left( \frac{\gamma }{\Gamma \left(
1-\alpha \right) }\right) ^{1/\alpha }Y_{\gamma }^{-1}k^{-1/\alpha }\text{
as }k\rightarrow \infty
\end{equation*}
and the ordered frequencies only decay algebraically fast with $k$. Species
with small frequency are long-tailed (there are many small size groups or
rare species in the Engen model, compared to the Ewens model).

In this model, $\phi _{\bullet }=\alpha \left( 1-\alpha \right) _{\bullet
-1}.$ Because $\phi _{1}=\alpha $ and $\phi _{m+1}=\phi _{m}\left( m-\alpha
\right) $, $m\geq 1$, it follows from (\ref{f2}, \ref{f2b}) that $\sigma
_{k+1}\left( \theta \right) =\left( \theta \alpha +k\right) \sigma
_{k}\left( \theta \right) -\theta \alpha \sigma _{k}^{\prime }\left( \theta
\right) .$ Thus, the Bell coefficients $B_{k,p}\left( \phi _{\bullet
}\right) ,$ appearing in (\ref{f9a}), again obey a simple $3-$term
recurrence 
\begin{equation*}
B_{k+1,p}\left( \phi _{\bullet }\right) =\alpha B_{k,p-1}\left( \phi
_{\bullet }\right) +\left( k-p\alpha \right) B_{k,p}\left( \phi _{\bullet
}\right) \text{.}
\end{equation*}
They constitute generalized Stirling numbers studied by \cite{Char}. It can
be checked that $\sigma _{k}\left( \theta \right) \notin ZR_{-}$.

This model is amenable to similar conclusions as the ones from the previous
example with recursion now given by 
\begin{equation*}
\mathbf{P}^{*}\left( P_{k+1}=p\right) =\frac{\sigma _{k}\left( \gamma
\right) }{\sigma _{k+1}\left( \gamma \right) }\left( \alpha \gamma \mathbf{P}%
^{*}\left( P_{k}=p-1\right) +\left( k-p\alpha \right) \mathbf{P}^{*}\left(
P_{k}=p\right) \right) .
\end{equation*}
Equation (\ref{f19}) with $\phi _{\bullet }=\alpha \left( 1-\alpha \right)
_{\bullet -1}$ is the Engen's extended negative binomial sampling formula 
\cite{Hosh1}. The particular case $\alpha =1/2$ is studied in \cite{Hosh2}.
The microcanonical distribution (\ref{f19a}) coincides when $\phi _{\bullet
}=\alpha \left( 1-\alpha \right) _{\bullet -1}$ with the one occurring in
the Pitman sampling formula (\cite{Hosh1}, Remark $3$).\newline

$\left( v\right) $ Let $\phi \left( x\right) $ solve the functional equation 
$\phi \left( x\right) =x\exp \phi \left( x\right) .$ Then $\phi \left(
x\right) =\sum_{m\geq 1}\frac{\phi _{m}}{m!}x^{m}$ with $\phi _{m}=m^{m-1}$
is the Cayley generating function appearing in the enumeration of rooted
labeled trees with $m$ nodes$.$ The convergence radius of this series is $%
x_{0}=e^{-1}$ with $\phi \left( x_{0}\right) =1$ and $\phi ^{\prime }\left(
x_{0}\right) =\infty .$ Clearly $\phi _{m}$ is log-convex, it is a Stieltjes
moment sequence and $\phi \in \mathcal{S}$. The associated Laplace exponent $%
\psi \left( x\right) =-\phi \left( -x\right) $ is the Lambert function.
Because $\psi \left( x\right) \sim \log x$ as $x\rightarrow \infty $, $%
\overline{\pi }\left( t\right) \sim -\log t$ as $t\rightarrow 0$ and $%
N_{+}\left( t\right) :=\#\left\{ k:\Delta _{\left( k\right) }\left( \gamma
\right) >t\right\} $ grows like $-\gamma \log t$ as $t\rightarrow 0$.
Besides, like in Example $\left( i\right) $%
\begin{equation*}
-\log S_{\left( k\right) ,\gamma }\sim k/\gamma \text{ as }k\rightarrow
\infty .
\end{equation*}
The partition function $Z_{\theta }\left( x\right) =\exp \theta \phi \left(
x\right) $ occurs in the enumeration of forests of Cayley trees. The Bell
coefficients are $B_{k,p}\left( \phi _{\bullet }\right) =\binom{k-1}{p-1}%
k^{k-p}$ (number of forests with $k$ nodes and $p$ trees) in accordance with
the global weights $\sigma _{k}\left( \theta \right) =\theta \left( k+\theta
\right) ^{k-1}.$ So $\sigma _{k}\left( \theta \right) \in ZR_{-}.$ Assuming $%
\theta $ known, the MLE estimator of $n$ in the finitely many species model
is given implicitly by $P=\widehat{n}\left( 1-\frac{\sigma _{k}\left( \left( 
\widehat{n}-1\right) \theta \right) }{\sigma _{k}\left( \widehat{n}\theta
\right) }\right) $, so here 
\begin{equation*}
P=\widehat{n}-\left( \widehat{n}-1\right) \left( 1-\frac{\theta }{k+\widehat{%
n}\theta }\right) ^{k-1}.
\end{equation*}
The MLE estimator of $\gamma $ in the infinitely many species model is given
by $P=\widehat{\gamma }\frac{\sigma _{k}^{\prime }\left( \widehat{\gamma }%
\right) }{\sigma _{k}\left( \widehat{\gamma }\right) }$, so here explicit 
\begin{equation*}
\widehat{\gamma }=\frac{k\left( P-1\right) }{k-P}.
\end{equation*}
The alternative (biased) estimator is $\widetilde{\gamma }=\frac{%
B_{k,P-1}\left( \phi _{\bullet }\right) }{B_{k,P}\left( \phi _{\bullet
}\right) }.$ Thus 
\begin{equation*}
\widetilde{\gamma }=\frac{k\left( P-1\right) }{k-P+1}=\frac{1}{\frac{1}{P-1}-%
\frac{1}{k}};
\end{equation*}
it is also explicit and very close to $\widehat{\gamma }$.\newline

$\left( vi\right) $ As a next example, let $\phi \left( x\right) $ solve the
functional equation $\phi \left( x\right) =xg\left( \phi \left( x\right)
\right) $ where $g\left( x\right) =\left( 1+bx\right) ^{a}$ with either $b>0$
and $a\geq 1$ or $a$ and $b$ both negative$.$ $\phi \left( x\right) $ is the
generating function appearing in the enumeration of rooted trees when the
generating function $g$ of the offspring is either (generalized) binomial or
negative binomial$.$ Then $\phi _{m}=\left( m-1\right) !\binom{am}{m-1}%
b^{m-1}$ are non-negative numbers. We conjecture that $\phi \in \mathcal{S}$%
. It holds \cite{HM} that $x_{0}=\left( ab\right) ^{-1}\left( 1-1/a\right)
^{a-1}$ with $\phi \left( x_{0}\right) =1/\left( b\left( a-1\right) \right) $
and $\phi ^{\prime }\left( x_{0}\right) =\infty .$ For this tree model first
discussed in \cite{Beres}, the Lagrange inversion formula gives \cite{AB} 
\begin{equation*}
B_{k,p}\left( \phi _{\bullet }\right) =\binom{k-1}{p-1}\left\{ ak\right\}
_{k-p}b^{k-p},
\end{equation*}
where $\left\{ a\right\} _{l}:=a\left( a-1\right) ...\left( a-l+1\right) $.
Recalling $\widetilde{\gamma }=\frac{B_{k,P-1}\left( \phi _{\bullet }\right) 
}{B_{k,P}\left( \phi _{\bullet }\right) },$ we get 
\begin{equation*}
\widetilde{\gamma }=\frac{b\left( P-1\right) }{k-P+1}\left( \left(
a-1\right) k+P\right) =\frac{b}{\frac{1}{P-1}-\frac{1}{k}}\left( a-1+\frac{P%
}{k}\right) ,
\end{equation*}
which is explicit. Again, would $1/k$ be close to $1/\left( P-1\right) $,
then $\widetilde{\gamma }$ would be estimated to be large. Would $%
a\rightarrow \pm \infty $, $b\rightarrow \pm 0$ while $ab\rightarrow 1$, we
recover the results just obtained for Cayley trees (consistently with $%
g\left( x\right) =\left( 1+bx\right) ^{a}\rightarrow e^{x}$). If $a=b=1$, we
recover Example $\left( iii\right) $ with $\alpha =1.$ When $k$ is large,
the minimum of $B_{k,p}^{2}\left( \phi _{\bullet }\right) /\left(
B_{k,p-1}\left( \phi _{\bullet }\right) B_{k,p+1}\left( \phi _{\bullet
}\right) \right) $ is attained when $p=\left[ \lambda k\right] $ for some $%
\lambda \in \left( 0,1\right) ,$ with value 
\begin{equation*}
\frac{\lambda }{1-\lambda }\frac{\left( 1-\lambda \right) k+1}{\lambda k-1}%
\frac{\left( a-1+\lambda \right) k+1}{\left( a-1+\lambda \right) k}\underset{%
k\rightarrow \infty }{\rightarrow }1
\end{equation*}
and the sequence $B_{k,p}\left( \phi _{\bullet }\right) $ is $p-$log-concave.%
\newline

$\left( vii\right) $ Let $\alpha >0$ and let $\phi \left( x\right)
=\sum_{m\geq 1}m^{-\alpha }x^{m}$ be the polylog function. The convergence
radius of this series is $x_{0}=1$ with $\phi \left( x_{0}\right) <\infty $
iff $\alpha >1$ and $\phi ^{\prime }\left( x_{0}\right) <\infty $ iff $%
\alpha >2.$ $\phi \left( x\right) $ is defined for $x<x_{0}$ and $\phi
\left( x\right) \rightarrow -\infty $ as $x\rightarrow -\infty .$ We have $%
\phi _{m}=m!m^{-\alpha }$ and $\left( \phi _{m}\right) _{m\geq 1}$
constitutes a log-convex sequence because for all $m\geq 2,$ 
\begin{eqnarray*}
\phi _{m+1}\phi _{m-1} &=&\left( m+1\right) !\left( m-1\right) !\left(
m^{2}-1\right) ^{-\alpha } \\
&>&\left( m+1\right) !\left( m-1\right) !m^{-2\alpha }>m!^{2}m^{-2\alpha
}=\phi _{m}^{2}.
\end{eqnarray*}
The sequence $\phi _{m}$ satisfies Carleman's condition $\sum_{m\geq 1}\phi
_{m}^{-1/\left( 2m\right) }=\infty $. Thus $\phi \in \mathcal{S}$\emph{\ }%
and $\psi \left( x\right) =-\phi \left( -x\right) $, $x>-1,$ is the Laplace
exponent of some polylog subordinator with L\'{e}vy measure $\pi $. Because $%
\phi \left( x\right) \sim -\left[ \log (-x)\right] ^{\alpha }/\Gamma \left(
1+\alpha \right) $ as $x\rightarrow -\infty $, \cite{CG}, $-\phi \left(
-x\right) =:\psi \left( x\right) \rightarrow \infty $ as $x\rightarrow
\infty $ and $\pi $ has infinite total mass. In this example, when $\alpha
>1 $, the weight of large clusters $\phi _{m}$ is smaller than in Example $%
\left( i\right) $ where $\phi _{m}=\left( m-1\right) !$. When $\alpha >1$,
we therefore expect small clusters sizes to be enhanced as in Example $%
\left( iv\right) $, but to a lesser extent. Because indeed $\overline{\pi }%
\left( t\right) \sim \left[ -\log t\right] ^{\alpha }/\Gamma \left( 1+\alpha
\right) $ as $t\rightarrow 0$, $N_{+}\left( t\right) :=\#\left\{ k:\Delta
_{\left( k\right) }\left( \gamma \right) >t\right\} $ grows like $\gamma
\left[ -\log t\right] ^{\alpha }/\Gamma \left( 1+\alpha \right) $ as $%
t\rightarrow 0$. Besides, 
\begin{equation*}
-\log S_{\left( k\right) ,\gamma }\sim \left( \Gamma \left( 1+\alpha \right)
/\gamma \right) ^{1/\alpha }k^{1/\alpha }\text{ as }k\rightarrow \infty
\end{equation*}
and the ordered frequencies decay exponentially fast, but now with $%
k^{1/\alpha }$ (in a `stretched exponential' Weibull way).\newline

$\left( viii\right) $ As another example with $\phi \in \mathcal{S}$ but
with $\pi $ integrable, consider the Mittag-Leffler function $\phi \left(
x\right) =\sum_{m\geq 1}\frac{1}{\Gamma \left( 1+m\alpha \right) }x^{m},$
where $\alpha \in \left( 0,1\right) .$ We have $\psi \left( x\right) :=-\phi
\left( -x\right) =:1-\varphi \left( x\right) $ where 
\begin{equation*}
\varphi \left( x\right) :=\sum_{m\geq 0}\frac{1}{\Gamma \left( 1+m\alpha
\right) }\left( -x\right) ^{m}.
\end{equation*}
$\varphi \left( x\right) $ is the Mittag-Leffler LST of the random variable $%
S_{\alpha }^{-\alpha }$ where $S_{\alpha }$ is an $\alpha -$stable random
variable with LST $\mathbf{E}\left( e^{-xS_{\alpha }}\right) =e^{-x^{\alpha
}},$ \cite{Po}. Here $\phi _{\bullet }=\frac{\Gamma \left( 1+\bullet \right) 
}{\Gamma \left( 1+\alpha \bullet \right) }$ and because of the latter link
with the Mittag-Leffler LST, the $\phi _{\bullet }$ sequence is log-convex
and $\phi \in \mathcal{S}$. For this model, the discrete abundance $\xi $ is
thus a Poisson sum of discrete Mittag-Leffler increments $\delta $ with 
\begin{equation*}
\mathbf{P}\left( \delta =m\right) =\frac{1}{\Gamma \left( 1+m\alpha \right) }%
\frac{x^{m}}{\phi \left( x\right) },\text{ }m\geq 1.
\end{equation*}
In the bias sampling from a random partition point of view, the
L\'{e}vy-measure corresponding to $Y_{\gamma }$ is $\pi \left( dt\right)
=f_{\alpha }\left( t\right) dt$ where $f_{\alpha }\left( t\right) $ is the
density of $S_{\alpha }^{-\alpha }.$ The Laplace exponent of $Y_{\gamma }$
is $\psi \left( x\right) =-\phi \left( -x\right) .$ Because $\pi $ is
integrable with mass $1$, $Y_{\gamma }$ is a subordinator in the compound
Poisson class (a Poisson$\left( \gamma \right) $ sum of iid positive jumps
with Mittag-Leffler density $f_{\alpha }\left( t\right) $). In the
Mittag-Leffler case, the bias sampling is again from a finite random
partition of unity, as in Example $\left( iii\right) $. Note that as $\alpha
\rightarrow 0$, $\phi \left( x\right) \sim \left( 1-x\right) ^{-1}-1$ (which
is a particular case of $\left( iii\right) $) whereas when $\alpha
\rightarrow 1$, $\phi \left( x\right) \sim e^{x}-1$ which is the Bell model,
also in the $\mathcal{S}$ class$.$\newline

$\left( ix\right) $ Let $\phi \left( x\right) $ solve the functional
equation $\phi \left( x\right) =xg\left( \phi \left( x\right) \right) $
where $g\left( x\right) =1+x^{2}/2.$ Then $\phi \left( x\right) =\left( 1-%
\sqrt{1-2x^{2}}\right) /x$ is the generating function appearing in the
enumeration of rooted binary labeled trees$.$ Only the odd $\phi _{m}$'s are
non-zero. The convergence radius of this series is $x_{0}=1/\sqrt{2}$ with $%
\phi \left( x_{0}\right) =\sqrt{2}$ and $\phi ^{\prime }\left( x_{0}\right)
=\infty .$ Clearly $\phi \notin \mathcal{S}$ because $\phi $ is only defined
on $\left| x\right| \leq x_{0}$, so not absolutely monotone on $\left(
-\infty ,x_{0}\right) $.

\section{A new Engen-like example}

We end up giving a new example of $\xi $ sharing some common issues with the
Engen's model.\newline

\textbf{Preliminaries.} Previously, let us start with a general fact. Let $%
\phi ^{\star }\left( x\right) $ be some `local' generating function with
non-negative coefficients $\phi _{m}^{\star }$. Define $Z_{1}^{\star }\left(
x\right) =\exp \phi ^{\star }\left( x\right) $, together with $\sigma
_{k}^{\star }\left( \theta \right) ,$ the Bell polynomials associated to $%
\phi ^{\star }\left( x\right) $: $Z_{1}^{\star }\left( x\right) ^{\theta
}=:1+\sum_{k\geq 1}\frac{\sigma _{k}^{\star }\left( \theta \right) }{k!}%
x^{k}.$ Define now the new generating functions 
\begin{equation*}
\phi \left( x\right) =xZ_{1}^{\star }\left( x\right) \text{ and }Z_{\theta
}\left( x\right) =\exp \left( \theta \phi \left( x\right) \right) .
\end{equation*}
The Taylor coefficients of $\phi $ are: $\phi _{m}=m\sigma _{m-1}^{\star
}\left( 1\right) .$ The Bell polynomials now associated to $\phi \left(
x\right) $ are: $Z_{\theta }\left( x\right) =1+\sum_{k\geq 1}\frac{\sigma
_{k}\left( \theta \right) }{k!}x^{k}$, with 
\begin{equation*}
\sigma _{k}\left( \theta \right) =\sum_{p=1}^{k}B_{k,p}\left( \bullet \sigma
_{\bullet -1}^{\star }\left( 1\right) \right) \theta ^{p}.
\end{equation*}
Because $\sigma _{k}^{\star }\left( \theta \right) $ are binomial
convolution polynomials, the following identity holds, \cite{AB} 
\begin{equation}
B_{k,p}\left( \bullet \sigma _{\bullet -1}^{\star }\left( 1\right) \right) =%
\binom{k}{p}\sigma _{k-p}^{\star }\left( p\right) .  \label{N1}
\end{equation}
Three simple examples are:

- $\phi ^{\star }\left( x\right) =\alpha x,$ $\alpha >0.$ Then $\sigma
_{k}^{\star }\left( \theta \right) =\alpha ^{k}\theta ^{k}$ leading to: $%
B_{k,p}\left( \bullet \alpha ^{\bullet -1}\right) =\binom{k}{p}\left( \alpha
p\right) ^{k-p}.$

- $\phi ^{\star }\left( x\right) =e^{\alpha x}-1,$ $\alpha >0.$ Then $\sigma
_{k}^{\star }\left( \theta \right) =\alpha ^{k}\sum_{p=1}^{k}S_{k,p}\theta
^{p}$ (where $S_{k,p}$ are the second kind Stirling numbers), leading to: $%
B_{k,p}\left( \alpha ^{\bullet -1}B_{\bullet -1}\right) =\binom{k}{p}\alpha
^{k-p}\sum_{q=1}^{k-p}S_{k-p,q}p^{q}$ where $B_{k}=\sum_{p=1}^{k}S_{k,p}$
are the Bell numbers$.$

- $\phi ^{\star }\left( x\right) $ solves $\phi ^{\star }\left( x\right)
=x\exp \left( \alpha \phi ^{\star }\left( x\right) \right) ,$ $\alpha >0.$
Then $\sigma _{k}^{\star }\left( \theta \right) =\sum_{p=1}^{k}B_{k,p}\left(
\phi _{\bullet }^{\star }\right) \theta ^{p}$ with $B_{k,p}\left( \phi
_{\bullet }^{\star }\right) =\binom{k-1}{p-1}\left( \alpha k\right) ^{k-p},$
leading to 
\begin{equation*}
\sigma _{k}^{\star }\left( \theta \right) =\theta \left( \theta +\alpha
k\right) ^{k-1}.
\end{equation*}
We conclude that, with $\phi _{\bullet }=\bullet \left( 1+\alpha \left(
\bullet -1\right) \right) ^{\bullet -2}$ 
\begin{equation*}
B_{k,p}\left( \phi _{\bullet }\right) =\binom{k}{p}p\left( p+\alpha \left(
k-p\right) \right) ^{k-p-1}.
\end{equation*}
If $\alpha =1,$ $\phi _{\bullet }=\bullet \left( 1+\alpha \left( \bullet
-1\right) \right) ^{\bullet -2}=\bullet ^{\bullet -1}$ and we recover $%
B_{k,p}\left( \bullet ^{\bullet -1}\right) =\binom{k}{p}pk^{k-p-1}=\binom{k-1%
}{p-1}k^{k-p}.$\newline

\textbf{The example.}

Let $\phi ^{\star }\left( x\right) =-\alpha \log \left( 1-x\right) ,$ $%
\alpha >0.$ Then $\sigma _{k}^{\star }\left( \theta \right) =\left( \alpha
\theta \right) _{k}.$ Looking at $\phi \left( x\right) =x\exp \phi ^{\star
}\left( x\right) $ and 
\begin{equation*}
Z_{\theta }\left( x\right) =\exp \left( \theta \phi \left( x\right) \right)
=e^{\theta x\left( 1-x\right) ^{-\alpha }},
\end{equation*}
with $\phi _{\bullet }=\bullet \left( \alpha \right) _{\bullet -1}$, we get $%
\sigma _{k}\left( \theta \right) =\sum_{p=1}^{k}B_{k,p}\left( \phi _{\bullet
}\right) \theta ^{p}$ where 
\begin{equation}
B_{k,p}\left( \bullet \left( \alpha \right) _{\bullet -1}\right) =\binom{k}{p%
}\left( \alpha p\right) _{k-p}.  \label{N2}
\end{equation}

\begin{proposition}
The new model $\phi \left( x\right) =x\left( 1-x\right) ^{-\alpha }\in 
\mathcal{S}$ iff $\alpha \in \left[ 0,1\right] .$
\end{proposition}

\textbf{Proof:} First, the convergence radius of $\phi $ is $x_{0}=1.$

We have $\phi ^{\prime }\left( x\right) =\left( 1-x\right) ^{-\left( \alpha
+1\right) }\left( 1-x\left( 1-\alpha \right) \right) $ and $\phi ^{\prime
}>0 $ for all $x<x_{0}$ only if $\alpha \in \left[ 0,1\right] .$ Let then $%
\alpha \in \left[ 0,1\right] $. Then $\phi ^{\left( k\right) }\left(
x\right) =\left( 1-x\right) ^{-\left( \alpha +k\right) }\left(
a_{k}-xb_{k}\right) $ and suppose both $a_{k}$ and $b_{k}$ are positive with 
$a_{k}/b_{k}>1$ in such a way that $\phi ^{\left( k\right) }>0$ for all $%
x<x_{0}.$ Then 
\begin{equation*}
\phi ^{\left( k+1\right) }\left( x\right) =\left( 1-x\right) ^{-\left(
\alpha +k+1\right) }\left( \left( \alpha +k\right) a_{k}-xb_{k}\left( \alpha
+k-1\right) \right)
\end{equation*}
with $a_{k+1}=\left( \alpha +k\right) a_{k}$ and $b_{k+1}=b_{k}\left( \alpha
+k-1\right) $. Both $a_{k+1}$ and $b_{k+1}$ are positive with $%
a_{k+1}/b_{k+1}>a_{k}/b_{k}>1.$ So $\phi ^{\left( k+1\right) }>0$ for all $%
x<x_{0}.$ $\diamond $

\begin{corollary}
When $\alpha \in \left( 0,1\right) $, in the infinitely many species
context, sampling from a discrete abundance model $\xi $ built on $\phi
\left( x\right) =x\left( 1-x\right) ^{-\alpha }$ interprets as bias sampling
from a random partition of unity $\mathbf{S}_{\infty }\left( \gamma \right) $
with ordered frequencies decaying algebraically fast with $k$. The Laplace
exponent associated to $Y_{\gamma }$ is $\psi \left( x\right) =-\phi \left(
-x\right) =x\left( 1+x\right) ^{-\alpha }$, $x>-1.$ The estimator $%
\widetilde{\gamma }$ of the biodiversity parameter $\gamma $ is explicitly
given by 
\begin{equation}
\widetilde{\gamma }=\frac{P}{k-P+1}\frac{\left( \alpha \left( P-1\right)
\right) _{k-P+1}}{\left( \alpha P\right) _{k-P}}.  \label{N3}
\end{equation}
\end{corollary}

\textbf{Proof:} Clearly $\psi \left( x\right) \sim x^{1-\alpha }\rightarrow
\infty $ as $x\rightarrow \infty $ and the corresponding L\'{e}vy measure $%
\pi $ has infinite mass.

We have $\overline{\pi }\left( t\right) \sim t^{-\left( 1-\alpha \right)
}\rightarrow \infty $ as $t\rightarrow 0$ so that $N_{+}\left( t\right)
:=\#\left\{ k:\Delta _{\left( k\right) }\left( \gamma \right) >t\right\} $
grows like $\gamma t^{-\left( 1-\alpha \right) }$ as $t\rightarrow 0$ and 
\begin{equation*}
S_{\left( k\right) ,\gamma }\sim Y_{\gamma }^{-1}\left( k/\gamma \right)
^{-1/\left( 1-\alpha \right) }\text{ as }k\rightarrow \infty .
\end{equation*}
Like in the Engen model, the ordered frequencies decay algebraically fast
with $k$.

The expression of $\widetilde{\gamma }$ in (\ref{N3}) follows from (\ref{N2}%
). $\diamond $\newline

When both $k$ and $P$ are large, together with $k-\left( 1-\alpha \right) P$%
, using a simple asymptotic form for (\ref{N2}) 
\begin{equation*}
\widetilde{\gamma }\sim \frac{P\left( k-\left( 1-\alpha \right) P\right) }{%
k-P+1}\left( 1+\frac{\alpha +k-P}{\alpha \left( P-1\right) }\right)
^{-\alpha }.
\end{equation*}
\textbf{Acknowledgments:} T.H. acknowledges partial support from the ANR
Mod\'{e}lisation Al\'{e}atoire en \'{E}cologie, G\'{e}n\'{e}tique et
\'{E}volution (ANR-Man\`{e}ge- 09-BLAN-0215 project) and from the labex
MME-DII (Mod\`{e}les Math\'{e}matiques et \'{E}conomiques de la Dynamique,
de l' Incertitude et des Interactions). Part of this work was done while
S.M. was visiting Professor at the University of Cergy-Pontoise. Both
authors thank support from Basal CONICYT project PFB-03. The authors are
indebted to their Referees and the Editor in Charge for suggesting
improvements and correcting some mistakes appearing in a former draft.%
\newline


\begin{thebibliography}{99}
\bibitem{AB}  Abbas, M.; Bouroubi, S. On new identities for Bell's
polynomials. Discrete Mathematics, Volume 293, Issues 13, Pages 5-10 (2005).

\bibitem{BDN}  Bahls, P.; Devitt-Ryder, R.; Nguyen, T. On the location of
roots of logaritmically concave polynomials. Preprint available at
http://facstaff.unca.edu/ pbahls/papers/ BahlsDevittRyderNguyenV2.pdf (2010).

\bibitem{Beres}  Berestycki, N.; Pitman, J. Gibbs distributions for random
partitions generated by a fragmentation process. Journal of Statistical
Physics, Volume 127, Number 2, 381-418 (2007).

\bibitem{Berg}  Berg, C.; Christensen, J. P. R.; Ressel, P. Harmonic
analysis on semigroups. Theory of positive definite and related functions.
Graduate Texts in Mathematics, 100. Springer-Verlag, New York, (1984).

\bibitem{Bern}  Bernstein, S. Sur les fonctions absolument monotones. Acta
Math. 52, no. 1, 1-66 (1929).

\bibitem{Bertoin}  Bertoin J., L\'{e}vy processes. Cambridge University
Press, Cambridge, (1996).

\bibitem{BM}  Blackwell, D.; MacQueen, J.B. Ferguson distributions via
P\'{o}lya urn schemes. The Annals of Statistics, 1, 353--355 (1973).

\bibitem{Bunge}  Bunge, J.; Fitzpatrick, M. Estimating the Number of
Species: A Review. Journal of the American Statistical Association, Vol. 88,
No. March 1998, pp. 364-37 (1998).

\bibitem{Char}  Charalambides, Ch. A.; Singh, J. A review of the Stirling
numbers, their generalizations and statistical applications. Comm. Statist.
Theory Methods, 17, no. 8 (1988).

\bibitem{Comtet}  Comtet, L. Analyse combinatoire. Tomes 1 et 2. Presses
Universitaires de France, Paris, (1970).

\bibitem{CG}  Costin, O.; Garoufalidis, S. Resurgence of the fractional
polylogarithms. Math. Res. Lett. 16, no. 5, 817-826 (2009).

\bibitem{Dar}  Darroch, J. N. On the distribution of the number of successes
in independent trials. Ann. Math. Statist. 35\textbf{,} 1317-1321 (1964).

\bibitem{DP}  Davenport, H.; P\'{o}lya, G. On the product of two power
series. Canadian J. Math.\textbf{,} no 1, 1-5 (1949).

\bibitem{Donelly}  Donnelly, P. Partition structures, P\`{o}lya urns, the
Ewens sampling formula and the age of alleles. Theoretical Population
Biology, 30, 271-288 (1986).

\bibitem{Eng}  Engen, S. On species frequency models. Biometrika\textbf{,}
no 61, 263-270 (1974).

\bibitem{Engen}  Engen, S. Stochastic abundance models. Monographs on
Applied Probability and Statistics, Chapman and Hall, London, (1978).

\bibitem{Ew}  Ewens, W.J. Some remarks on the law of succession. Athens
Conference on Applied Probability and Time Series Analysis (1995), Vol. 
\textbf{I}, 229--244, Lecture Notes in Statistics, 114, Springer, New York
(1996).

\bibitem{Ewen}  Ewens, W.J. The sampling theory of selectively neutral
alleles. Theoretical Population Biology, 3\textbf{,} 87-112 (1972).

\bibitem{Ewens}  Ewens, W.J. Population genetics theory - the past and the
future. In: Mathematical and Statistical Developments of Evolutionary
Theory, S. Lessard Edt., Kluwer, Dordrecht, (1990).

\bibitem{Feng}  Feng, S. The Poisson-Dirichlet distribution and related
topics. Models and asymptotic behaviors. Probability and its Applications
(New York). Springer, Heidelberg, (2010).

\bibitem{FCW}  Fisher, R. A.; Corbet, A. S.; Williams, C. B. The relation
between the number of species and the number of individuals in a random
sample of an animal population. Journal of Animal Ecology, 12, 42-58 (1943).

\bibitem{GS}  Garibaldi, U.; Scalas, E. Finitary probabilistic methods in
econophysics. Cambridge University Press, Cambridge (2010).

\bibitem{GP}  Gnedin, A.; Pitman, J. Exchangeable Gibbs partitions and
Stirling triangles. (English, Russian summary) Zap. Nauchn. Sem.
S.-Peterburg. Otdel. Mat. Inst. Steklov. (POMI) 325 (2005), Teor. Predst.
Din. Sist. Komb. i Algoritm. Metody. 12, 83--102, 244--245; translation in
J. Math. Sci. (N. Y.) 138 (2006), no. 3.

\bibitem{HLP}  Hardy, G. H.; Littlewood, J. E.; P\'{o}lya, G. Inequalities.
2d ed. Cambridge, at the University Press (1952).

\bibitem{Ho}  Ho, M.W.; James, L.; Lau, J.W. Gibbs Partitions (EPPF's)
Derived From a Stable Subordinator are Fox H and Meijer G Transforms.
http://arxiv.org/abs/0708.0619, (2007).

\bibitem{Holst}  Holst, L. The Poisson-Dirichlet distribution and its
relatives revisited. available at:
http://www.math.kth.se/matstat/fofu/reports/PoiDir.pdf (2001).

\bibitem{Hosh1}  Hoshino, N. Engen's extended negative binomial model
revisited. Ann. Inst. Statist. Math., 57, No. 2, 369--387 (2005).

\bibitem{Hosh2}  Hoshino, N. Random clustering based on the conditional
inverse Gaussian-Poisson distribution. J. Japan Statist. Soc., 33, No. 1,
105--117 (2003).

\bibitem{Hub}  Hubbell, S. P. The neutral theory of biodiversity and
biogeography and Stephen Jay Gould. Paleobiology 31, 122-123 (2005).

\bibitem{Hui}  Huillet, T. Unordered and ordered sample from Dirichlet
distribution. Ann. Inst. Statist. Math., Vol 57, Issue 3, 597-616 (2005).

\bibitem{HM}  Huillet, T.; M\"{o}hle, M. Asymptotics of symmetric compound
Poisson population models. Submitted to Combinatorics, Probability and
Computing, Special issue dedicated to the memory of Philippe Flajolet,
Preprint available at hal-00730734 (2012).

\bibitem{HM1}  Huillet, T.; M\"{o}hle, M. Correction on `Population genetics
models with skewed fertilities: a forward and backward analysis'. Stoch.
Models 28, no. 3, 527-532, (2012).

\bibitem{Keener}  Keener, R; Rothman, E.; Starr, N. Distributions on
partitions. Ann. Statist. 15, no. 4, 1466-1481 (1987).

\bibitem{Kingm}  Kingman, J.F.C. Random discrete distributions. Journal of
the Royal Statistical Society. Series B, 37, 1--22 (1975).

\bibitem{King}  Kingman, J. F. C. Mathematics of genetic diversity. CBMS-NSF
Regional Conference Series in Applied Mathematics, 34. Society for
Industrial and Applied Mathematics (SIAM), Philadelphia, Pa. (1980).

\bibitem{Kingman}  Kingman, J.F.C. Poisson processes. Clarendon Press,
Oxford (1993).

\bibitem{Kol}  Kolchin, V. F. Random mappings. Translated from the Russian.
With a foreword by S. R. S. Varadhan. Translation Series in Mathematics and
Engineering. Optimization Software, Inc., Publications Division, New York
(1986).

\bibitem{Kol2}  Kolchin, V. F. Random graphs. Encyclopedia of Mathematics
and its Applications, 53. Cambridge University Press, Cambridge, (1999).

\bibitem{Moh}  M\"{o}hle, M. The concept of duality and applications to
Markov processes arising in neutral population genetics models. Bernoulli 5
(1999), no. 5, 761--777.

\bibitem{Neveu}  Neveu, J. Processus ponctuels. \'{E}cole d' \'{E}t\'{e} de
Probabilit\'{e}s de Saint-Flour, VI 1976, pp. 249-445. Lecture Notes in
Math., Vol. 598, Springer-Verlag, Berlin (1977).

\bibitem{Pit1}  Pitman, J. Random discrete distributions invariant under
size-biased permutation. Advances in Applied Probability,\emph{\ }28,
525-539 (1996).

\bibitem{Pit3}  Pitman, J. Exchangeable and partially exchangeable random
partitions. Probability Theory and Related Fields, 102, 145-158 (1995).

\bibitem{PitYor}  Pitman, J.; Yor, M. The two parameter Poisson-Dirichlet
distribution derived from a stable subordinator. Annals of Probability, 25,
855-900 (1997).

\bibitem{PitPK}  Pitman, J. Poisson-Kingman partitions. Statistics and
science: a Festschrift for Terry Speed, 134, IMS Lecture Notes Monogr. Ser.,
40, Inst. Math. Statist., Beachwood, OH, (2003).

\bibitem{PitCSP}  Pitman, J. Combinatorial stochastic processes. Lectures
from the $32$nd Summer School on Probability Theory held in Saint-Flour,
July 724, 2002. With a foreword by Jean Picard. Lecture Notes in
Mathematics, 1875. Springer-Verlag, Berlin, (2006).

\bibitem{Po}  Pollard, H. The completely monotonic character of the
Mittag-Leffler function $Ea\left( -x\right) $. Bull. Amer. Math. Soc. 54,
1115-1116 (1948).

\bibitem{IP}  Pr\"{u}nster, I. bibliography:
http://sites.carloalberto.org/pruenster/publications.html

\bibitem{Schill}  Schilling, R. L.; Song, R.; Vondracek, Z. Bernstein
functions. Theory and applications. de Gruyter Studies in Mathematics, 37.
Walter de Gruyter \& Co., Berlin, (2010).

\bibitem{scho}  Schoenberg, I. J. On the zeros of the generating functions
of multiply positive sequences and functions. Ann. of Math. (2), 62, 447-471
(1955).

\bibitem{Steu}  Steutel, F. W.; van Harn, K. Infinite divisibility of
probability distributions on the real line. Monographs and Textbooks in Pure
and Applied Mathematics, 259. Marcel Dekker, Inc., New York (2004).

\bibitem{Tavare}  Tavar\'{e}, S.; Ewens, W.J. Multivariate Ewens
distribution. Chapter 41 in Discrete Multivariate Distributions, N.L.
Johnson, S. Kotz and N. Balakrishnan Edts, Wiley, New York, 232-246 (1997).

\bibitem{Wat}  Watterson, G. A. The stationary distribution of the
infinitely-many neutral alleles diffusion model. J. Appl. Probability 13,
no. 4, 639-651, (1976).

\bibitem{Yang}  Yang, S.L. Some identities involving the binomial sequences.
Discrete Mathematics, Volume 308, Pages 51-58 (2008).
\end{thebibliography}
\end{document}